\documentclass{aa}  

\usepackage[breaklinks=true,colorlinks,citecolor=blue]{hyperref}
\usepackage{graphicx}
\usepackage{enumerate}
\usepackage{epsfig}
\usepackage{color}
\usepackage{txfonts}
\usepackage{natbib}
\usepackage{multirow}
\usepackage{ulem} 
\usepackage{amssymb}
\usepackage{xcolor}

\defcitealias{KhoperskovHESTIA-1}{Paper I}
\defcitealias{KhoperskovHESTIA-2}{Paper II}
\defcitealias{KhoperskovHESTIA-3}{Paper III}

\definecolor{ao}{rgb}{0.0, 0.5, 0.0}

\newcommand{\kmps}{\rm km~s\ensuremath{^{-1} }\,}

\newcommand{\Msun}{M\ensuremath{_\odot}}

\newcommand{\Msunpc}{\rm M\ensuremath{_\odot}~pc\ensuremath{^{-2} }\,}

\newcommand{\Gaia}{{\it Gaia}\,}

\newcommand{\FeH}{\ensuremath{\rm [Fe/H]}\,}

\begin{document} 

\title{The stellar halo in Local Group Hestia simulations I. \\ The in situ component and the effect of mergers}

\titlerunning{HESTIA simulations: in situ stellar haloes}

\author{Sergey Khoperskov$^1$\thanks{E-mail: sergey.khoperskov@gmail.com}, Ivan Minchev$^1$, Noam Libeskind$^{1,2}$, Misha Haywood$^3$, Paola Di Matteo$^3$, \\ Vasily Belokurov$^{4,5}$, Matthias Steinmetz$^1$,  Facundo A. Gomez$^{6,7}$, Robert J. J. Grand$^{8,9,10}$,  \\Yehuda Hoffman$^{11}$, Alexander Knebe$^{12,13,14}$,   Jenny G. Sorce$^{15.16,1}$, Martin Spaare$^{17,1}$, \\ Elmo Tempel$^{18,19}$, Mark Vogelsberger$^{20}$}

\authorrunning{Khoperskov et al.}

\institute{$^1$ Leibniz-Institut für Astrophysik Potsdam (AIP), An der Sternwarte 16, 14482 Potsdam, Germany\\
        $^2$ University of Lyon, UCB Lyon 1, CNRS/IN2P3, IUF, IP2I Lyon, France \\
        $^3$ GEPI, Observatoire de Paris, PSL Research University, CNRS, Place Jules Janssen, 92195 Meudon, France \\
        $^4$ Institute of Astronomy, Madingley Road, Cambridge CB3 0HA, UK \\
        $^5$ Center for Computational Astrophysics, Flatiron Institute, 162 5th Avenue, New York, NY 10010, USA \\
        $^6$ Instituto de Investigación Multidisciplinar en Ciencia y Tecnología, Universidad de La Serena, Raúl Bitrán 1305, La Serena, Chile \\
        $^7$ Departamento de Astronomía, Universidad de La Serena, Av. Juan Cisternas 1200 Norte, La Serena, Chile \\
        $^8$ Max-Planck-Institut für Astrophysik, Karl-Schwarzschild-Str 1, D-85748 Garching, Germany \\
        $^9$ Instituto de Astrofísica de Canarias, Calle Váa Láctea s/n, E-38205 La Laguna, Tenerife, Spain\\
        $^{10}$ Departamento de Astrofísica, Universidad de La Laguna, Av. del Astrofísico Francisco Sánchez s/n, E-38206 La Laguna, Tenerife, Spain\\
        $^{11}$ Racah Institute of Physics, Hebrew University, Jerusalem 91904, Israel \\
        $^{12}$ Departamento de Física Teórica, Módulo 15, Facultad de Ciencias, Universidad Autónoma de Madrid, E-28049 Madrid, Spain \\
        $^{13}$ Centro de Investigación Avanzada en Física Fundamental (CIAFF), Facultad de Ciencias, Universidad Autónoma de Madrid, E-28049 Madrid, Spain \\ 
        $^{14}$ International Centre for Radio Astronomy Research, University of Western Australia, 35 Stirling Highway, Crawley, Western Australia 6009, Australia \\
        $^{15}$ Univ. Lille, CNRS, Centrale Lille, UMR 9189 CRIStAL, F-59000 Lille, France\\
        $^{16}$ Universit\'e Paris-Saclay, CNRS, Institut d'Astrophysique Spatiale, 91405, Orsay, France\\
        $^{17}$ Institut für Physik und Astronomie, Universität Potsdam, Campus Golm, Haus 28, Karl-Liebknecht Straße 24-25, D-14476 Potsdam \\
        $^{18}$ Tartu Observatory, University of Tartu, Observatooriumi 1, 61602 Tõravere, Estonia \\
        $^{19}$ Estonian Academy of Sciences, Kohtu 6, 10130 Tallinn, Estonia \\
        $^{20}$ Department of Physics, Kavli Institute for Astrophysics and Space Research, Massachusetts Institute of Technology, Cambridge, MA 02139, USA
        }
        
\date{Received ; accepted }
 
\abstract{
Theory suggests that mergers play an important role in shaping galactic discs and stellar haloes, which was observationally confirmed in the Milky Way~(MW) thanks to \Gaia data. In this work, aiming to probe the contribution of mergers to the in situ stellar halo formation, we analyse six M31 and MW analogues from the HESTIA suite of cosmological hydrodynamical zoom-in simulations of the Local Group. We found that all the HESTIA galaxies experience between one to four mergers with stellar mass ratios between 0.2 and 1 relative to the host at the time of the merger. These significant mergers, with a single exception, happened $7-11$~Gyr ago. The overall impact of the most massive mergers in HESTIA is clearly seen as a sharp increase in the orbital eccentricity~(and a corresponding decrease in the rotational velocity $V_\phi$) of pre-existing disc stars of the main progenitor, thus nicely reproducing the Splash-, Plume-like feature that was discovered in the MW. We do find a correlation between mergers and close pericentric passages of massive satellites and bursts of the star formation in the in situ component. Massive mergers sharply increase the disc velocity dispersion of the in situ stars; however, the latest significant merger often heats up the disc up to the numbers when the contribution of the previous ones is less prominent in the age-velocity dispersion relation. In HESTIA galaxies, the in situ halo is an important component of the inner stellar halo where its fraction is about $30-40\%$, while in the outer parts it typically does not exceed $\approx 5\%$ beyond $15$~kpc from the galactic centre. The simulations suggest that this component of the stellar haloes continues to grow well after mergers conclude; however, the most significant contribution comes from stars that formed recently before the merger. The orbital analysis of the HESTIA galaxies suggests that wedges in $\rm R_{max}-Z_{max}$~(apocentre - maximum height from the mid-plane) space are mainly populated by the stars born in between significant mergers. 
}


\maketitle

\section{Introduction}\label{sec1::intro}
According to $\Lambda$ cold dark matter~(CDM) cosmology, galaxies acquire their mass via the continuous merging of smaller satellites~\citep{1991ApJ...379...52W,2005Natur.435..629S,2015MNRAS.446..521S} and accretion of gas from large-scale filaments~\citep{2005MNRAS.363....2K,2009Natur.457..451D}. Ancient episodes of gas accretion are hard to detect in local galaxies because of the dissipative nature of the ISM, while stellar merger remnants scattered across large spatial scales around galaxies, still could be seen as shells, streams, filaments and tidal tails~\citep{1983ApJ...274..534M, 2001Natur.412...49I, 2006ApJ...642L.137B, 2009Natur.461...66M, 2010AJ....140..962M, 2013ApJ...765...28A, 2015MNRAS.446..120D}. While being minor in terms of the mass budget at $\rm z=0$~\citep{2008ApJ...680..295B, 2011MNRAS.416.2903D, 2013ApJ...763..113D, 2019MNRAS.490.3426D, 2020MNRAS.492.3631M}, individual accretion events can cause transformation of host galaxies in the past and by contributing to the inner stellar halo~\citep{2008ApJ...680..295B, 2009ApJ...702.1058Z, 2010MNRAS.404.1711P, 2015ApJ...799..184P, 2015MNRAS.454.3185C, 2016ApJ...821....5D, 2018MNRAS.474.5300D}.

Various models suggest that classical bulges can be formed through mergers~\citep[see, e.g.][]{2005A&A...437...69B,2014MNRAS.440.2843C,2015ApJ...799..184P}. In less violent events, a number of different externally driven processes can result in the thickening of galactic discs, for example, heating of a pre-existing thin disc by minor mergers~\citep{1993ApJ...403...74Q,2008MNRAS.391.1806V, 2015ApJ...804L...9M}, accretion of disrupted satellites~\citep{2000AJ....119.2843C,2003ApJ...597...21A}, and formation of thick stellar components by gas-rich mergers~\citep{2004ApJ...612..894B,2005ApJ...630..298B}. In theory, disentangling the impact of various types of mergers can be done by comparing the eccentricity distributions for both in situ and accreted stellar populations~\citep[see, e.g.][]{2009MNRAS.400L..61S, 2011A&A...525L...3D, 2011A&A...530A..10Q}. However, one could expect that several mechanisms can contribute to the observed orbital composition of stellar populations even in a single galaxy~\citep[see, e.g.][]{2011MNRAS.413.2235W,2019MNRAS.482.3426M}, thus making it hard to connect the present-day characteristics with the past.

\begin{figure*}[t!]
\begin{center}
\includegraphics[width=1\hsize]{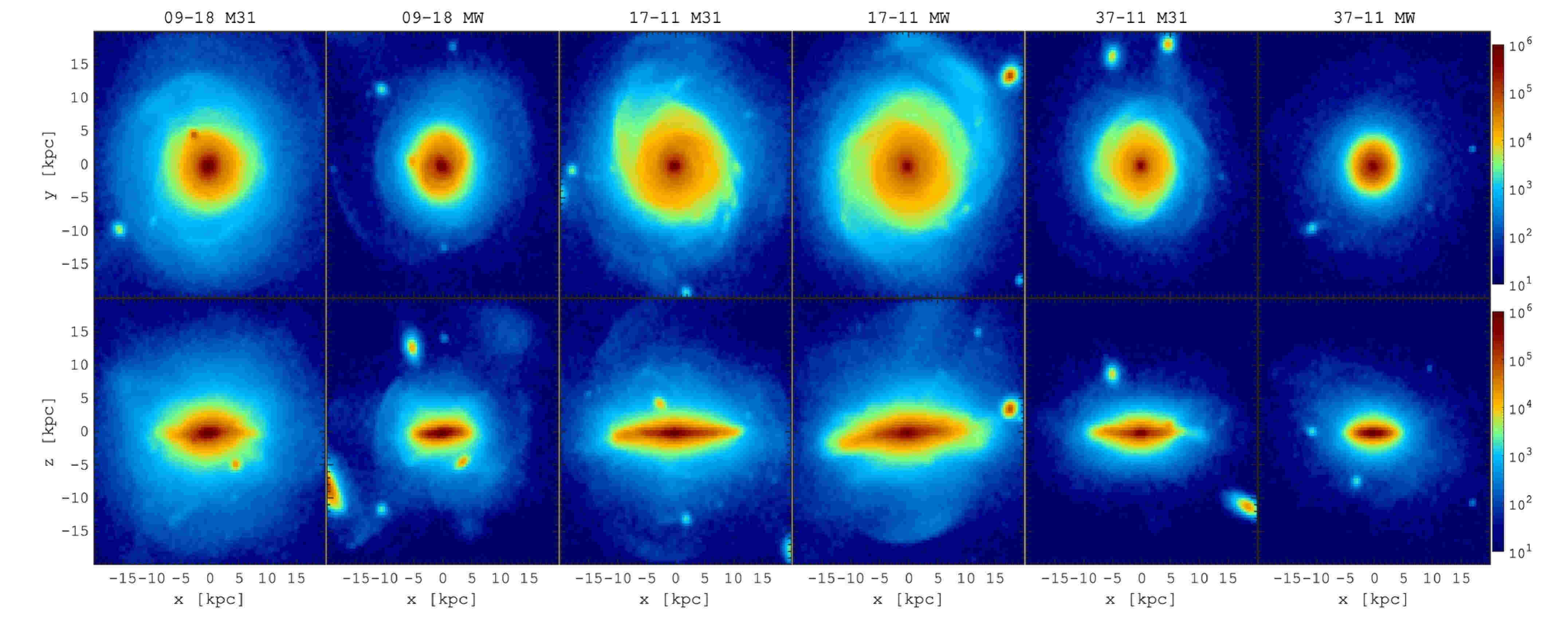}
\caption{Face-on (top) and edge-on (bottom) stellar density maps~($\Msunpc$) of the M31 and MW analogues in the HESTIA simulations. Disc components of the simulated galaxies have different extensions; however, all the galaxies show a presence of isolated stellar overdensities -- dwarf galaxies orbiting around the host galaxies -- and many smooth streams, shells, and tidal features contributing to the smooth halo component made of stars that formed ex situ and that accreted at different epochs. Mock HST images, gas distributions and a visualisation of the HESTIA Local Group are presented in \cite{2020MNRAS.498.2968L}. }\label{fig1::density_maps}
\end{center}
\end{figure*}

\begin{figure}[t!]
\begin{center}
\includegraphics[width=1\hsize]{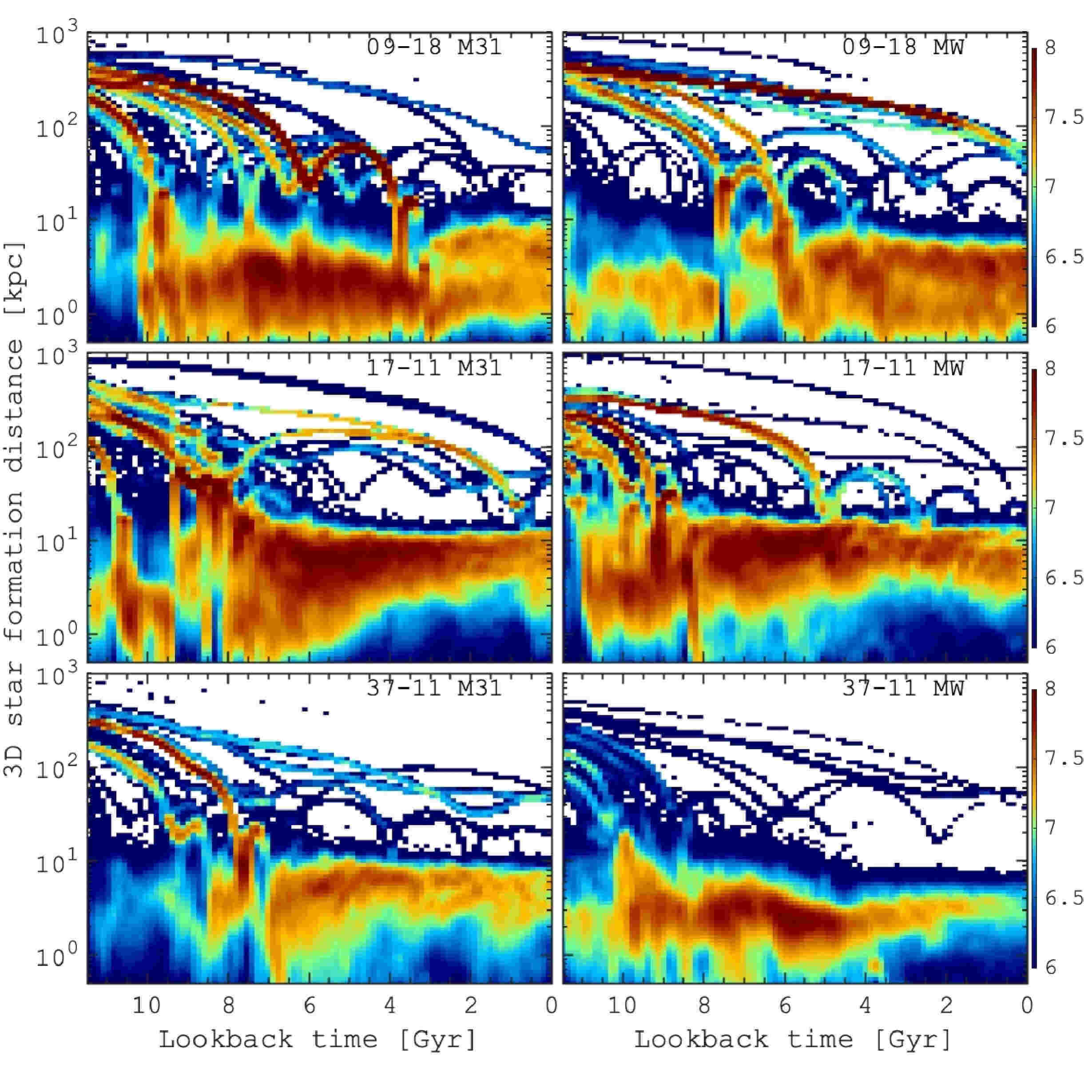}
\caption{Formation histories of the HESTIA simulations: formation distance of stars as a function of lookback time, colour-coded by the stellar mass~(in $\log_{10}$ units). The diagrams include all the in situ stars, merger remnants, and satellites that survived until $\rm z=0$. The HESTIA galaxies show a diverse assembly history with substantial early accretion and a different number of massive dwarfs in the halo at the present epoch.}\label{fig1::re1_time}
\end{center}
\end{figure}
\begin{figure*}[t!]
\begin{center}
\includegraphics[width=1\hsize]{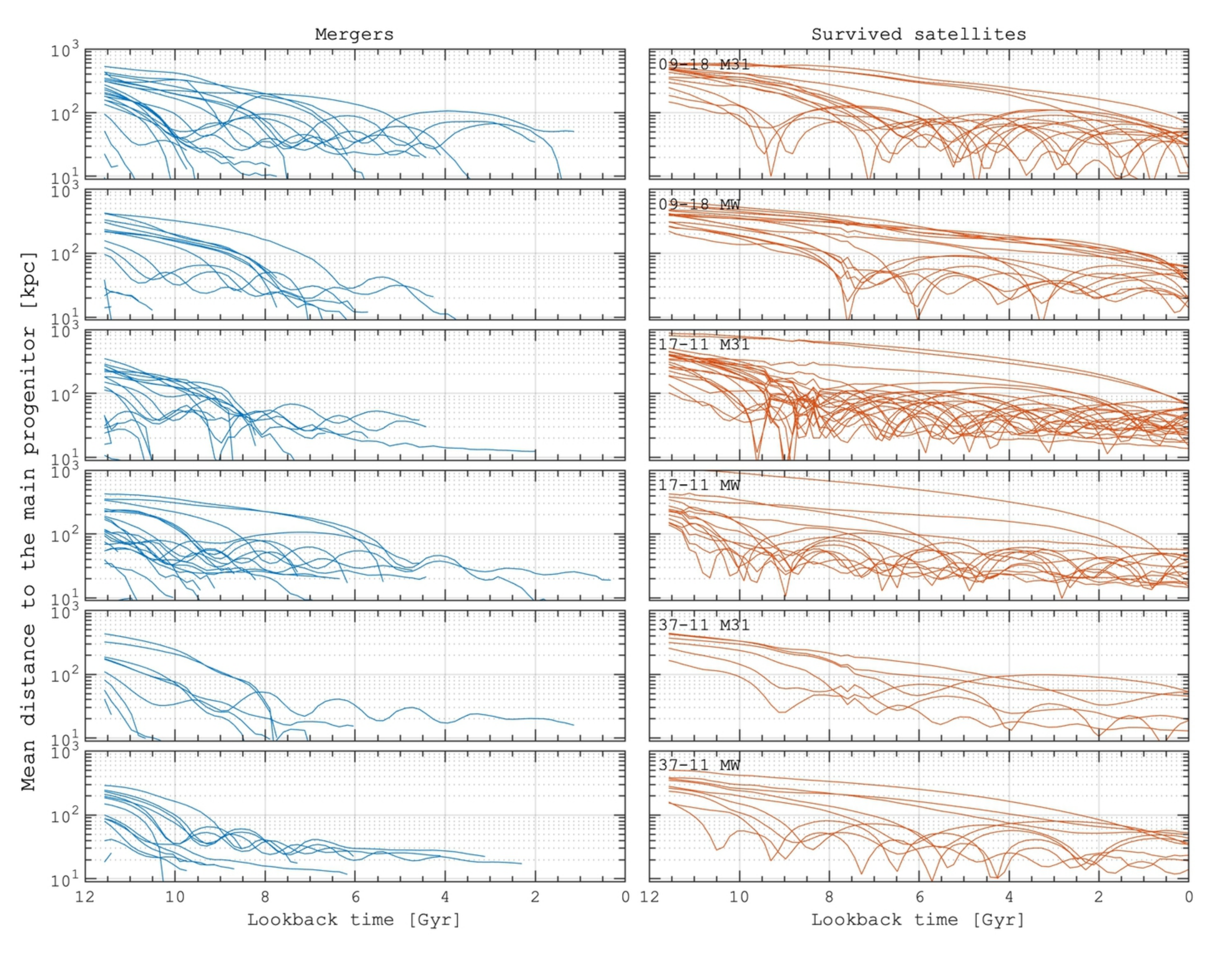}
\caption{Orbital decay for merged dwarf galaxies~(left) and satellites that survived~(right) for each M31 and MW analogue in the HESTIA simulations. The satellites that survived are the dwarf galaxies inside the virial radius of the main progenitor, which are still gravitationally bound to their own sub-halo at $\rm z=0$. Mergers correspond to the galaxies that, at some point in the evolution, become unbound and were disrupted inside the main progenitor. We note that some of these galaxies become unbound well before their mass centre reaches the centre of the main progenitor.}\label{fig1::sat_orbits}
\end{center}
\end{figure*}

\begin{figure*}[t!]
\begin{center}
\includegraphics[width=1\hsize]{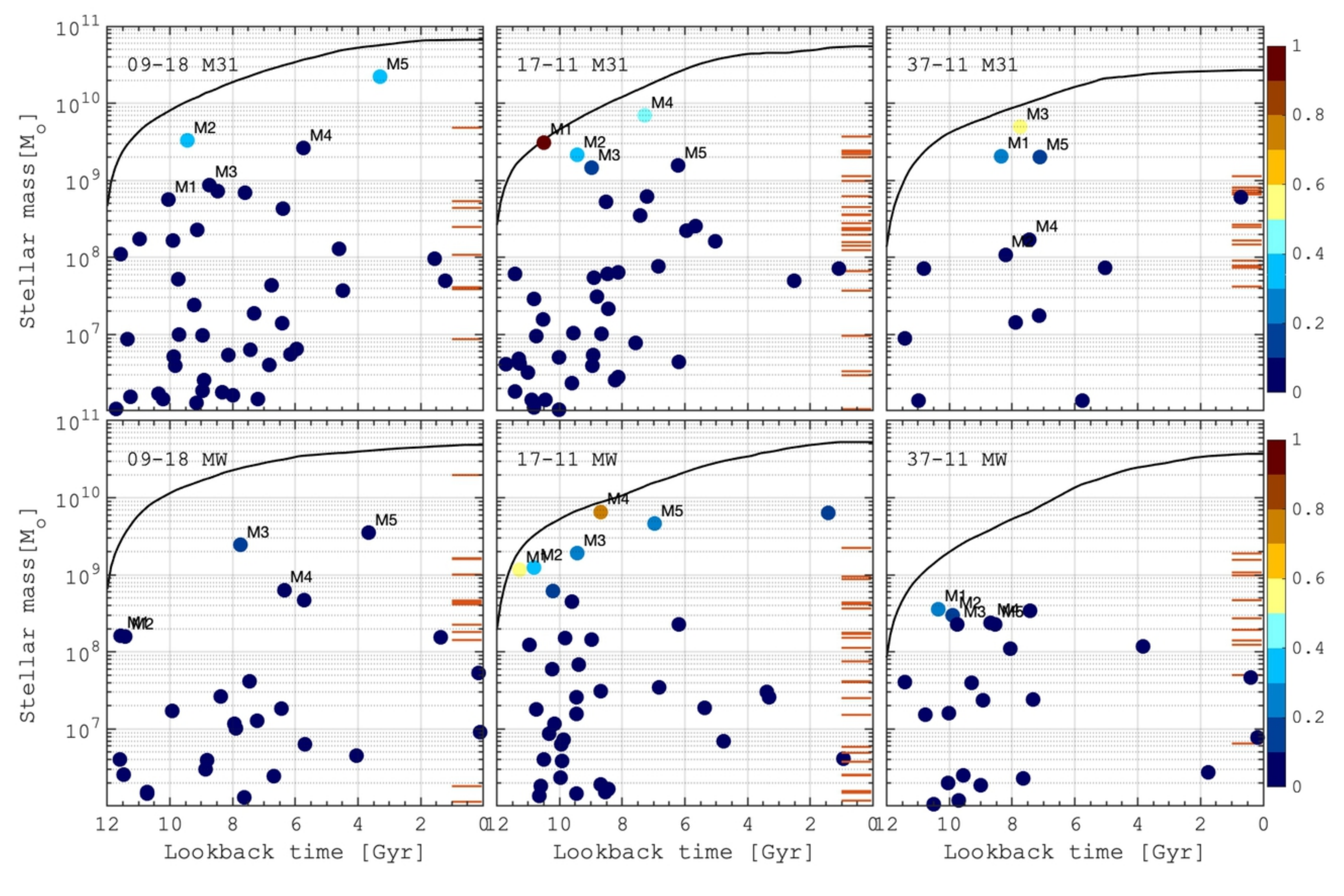}
\caption{Mergers history of the HESTIA galaxies. The in situ stellar mass growth for the six M31 and MW HESTIA galaxies is shown by the black line. Mergers are shown by the coloured circles representing the time of accretion and the stellar mass of a satellite at that time. The circles are colour-coded by the stellar mass ratio relative to the host~($\rm \mu_{*}$, colour bars are on the right). Red horizontal lines show the populations of dwarf galaxies that survive until $\rm z=0$. Five of the most significant mergers are marked as M1-M5 where the number increases from the earliest to the most recent ones. The HESTIA galaxies experienced between one to four massive merger events with the stellar mass ratio $>10\%$ relative to the host in situ component, while the other mergers are less significant in terms of stellar mass. The number of merger events significantly decreases starting from $\approx 8$~Gyr ago.}\label{fig1::mergers}
\end{center}
\end{figure*}

\begin{figure*}[t!]
\begin{center}
\includegraphics[width=1\hsize]{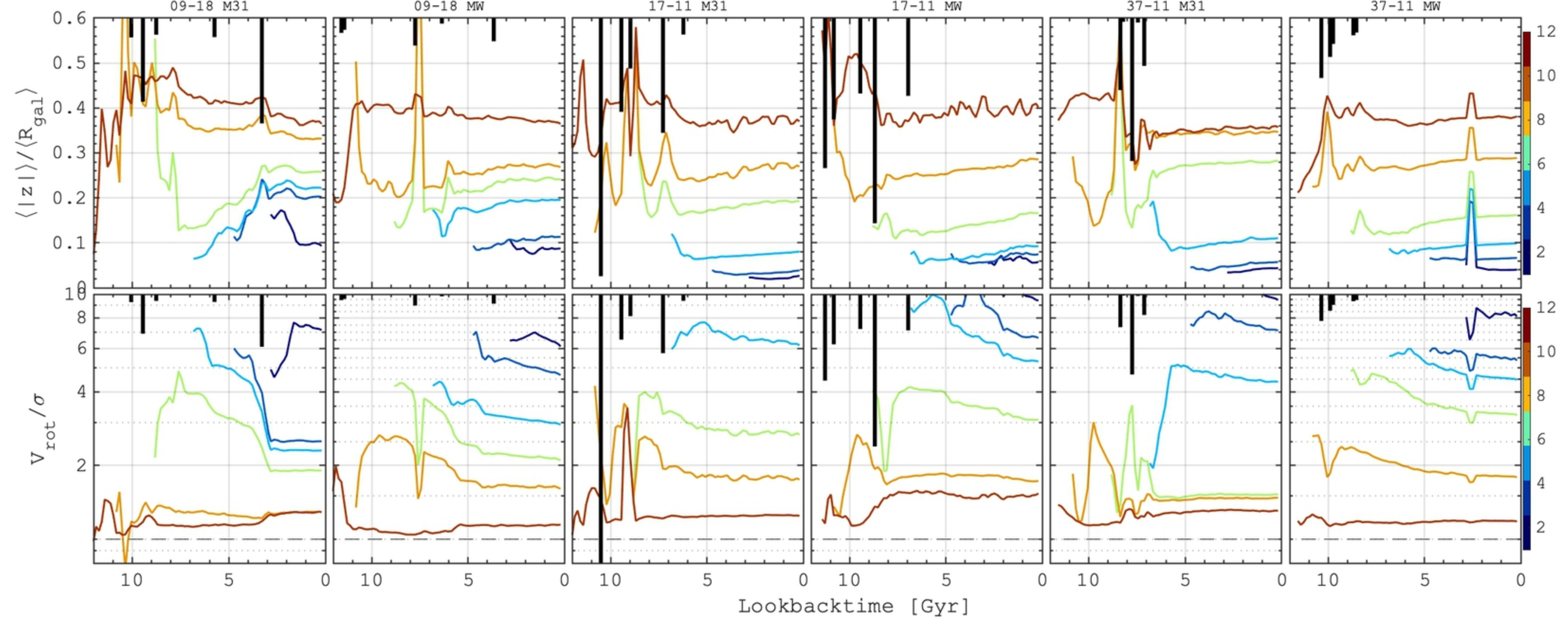}\caption{
Time evolution of the short-to-long-axis ratio and $V_\phi/\sigma$ relation of the in situ stellar populations. {\it Top:} Ratio between the mean height and the mean radial extension of the stellar component, calculated as the mean absolute values of the vertical, $\langle|z|\rangle$, and galactocentric radial, $\langle R_{gal}\rangle$, position of stars of a given age. {\it Bottom:} Ratio between the mean rotational velocity~($V_\phi$) and the radial velocity dispersion of stars in a given age range. Stars of different ages are shown by different colours, specified by the colour bar on the right, where each line starts from the oldest stars formed in a $2$~Gyr age-interval. The grey vertical lines highlight the five most significant mergers. The line length corresponds to the stellar mass ratio of the merger relative to the host at the time of accretion. The black dashed line in the bottom row shows $V_\phi/\sigma_R=1$. The figure suggests that in the HESTIA simulations even the oldest stars of M31 and MW galaxies were formed as flattened 'discy' components which however formed and remained dispersion-dominated populations. The impact of the mergers is seen in some cases, which, however, do not have a dominant impact on the shape and overall kinematics of the in situ stars. }\label{fig1::flat_or_round}
\end{center}
\end{figure*}

\begin{figure*}[t!]
\begin{center}
\includegraphics[width=1.0\hsize]{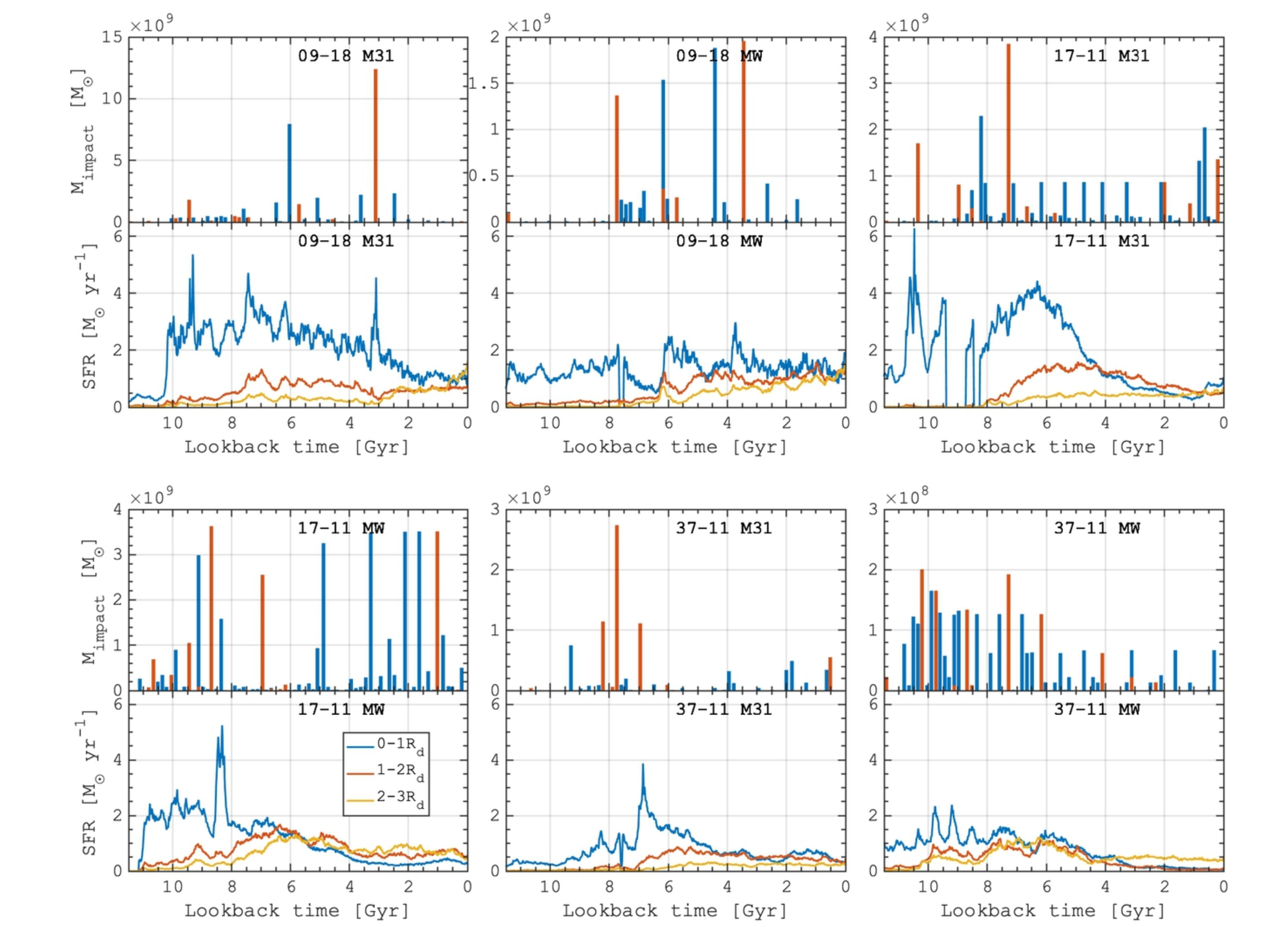}
\caption{Impact of the most significant mergers and close pericentric passages of massive satellites on the star formation history of M31 and MW HESTIA galaxies. For each galaxy, in the top panels we highlight the time of the close passages~($<10$ kpc from the host) and the stellar mass~($\rm M_{impact}$ at the time of the close passage) of the dwarf galaxies which either merged~(red) or survived at $\rm z=0$~(blue). The bottom panels show the star formation rate as a function of the lookback time for three galactocentric regions: $\rm  (0-1)R_d$~(blue), $\rm  (1-2)R_d$~(red), and $\rm (2-3)R_d$~(green), where $R_d $ is the disc scale length at $\rm z=0$. The star formation histories at different galactocentric radii are scaled by the mean star formation rate whose value is mentioned in each panel with the same colour. The star formation histories of the HESTIA galaxies show a number of short timescale peaks in addition to the early rise and slow decrease of the star formation rate. In most cases, the peaks of the star formation correlate with either massive mergers or with the pericentric passages of massive dwarf galaxies. However, not all of the close encounters correlate with the star formation bursts inside the host galaxies.}\label{fig1::sfr_encounters}
\end{center}
\end{figure*}

\begin{figure*}[t!]
\begin{center}
\includegraphics[width=1.0\hsize]{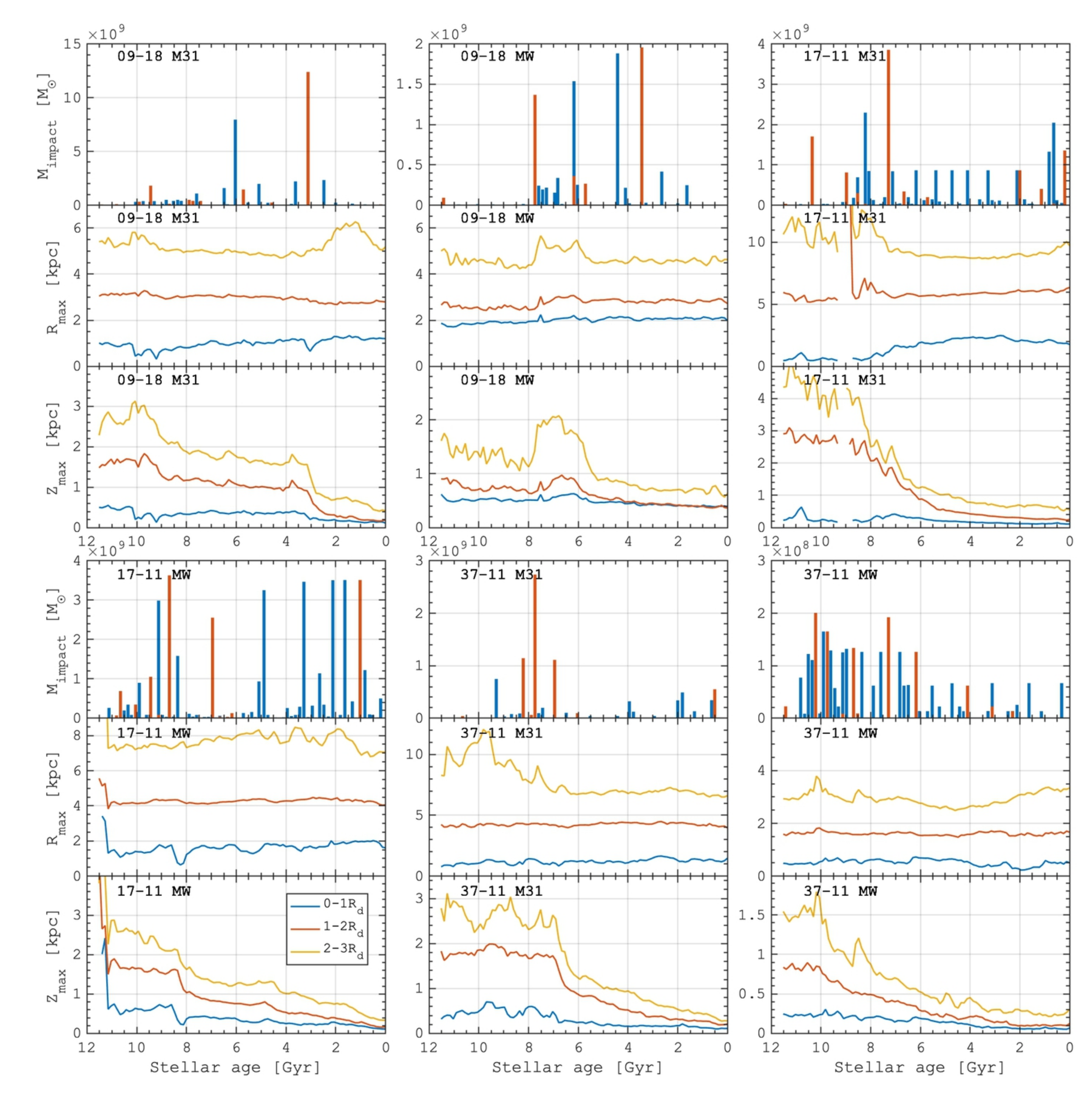}
\caption{Impact of the most significant mergers and close pericentric passages of massive satellites on the orbital parameters of the in situ stars in M31 and MW HESTIA galaxies. For each galaxy, in the top panels we highlight the time of the close passages~($<10$ kpc from the host) and the stellar mass~($\rm M_{impact}$ at the time of the close passage) of the dwarf galaxies which either merged~(red) or survived at $\rm z=0$~(blue). The two panels below show $Z_{max}$ and $Z_{max}/R_{max}$ as a function of the stellar age for three galactocentric regions: $\rm  (0-1)R_d$~(blue), $\rm  (1-2)R_d$~(red), and $\rm  (2-3)R_d$~(green), where $R_d $ is the disc scale length at $\rm z=0$.}\label{fig1::rmax_zmax_encounters}
\end{center}
\end{figure*}

\begin{figure*}[t!]
\begin{center}
\includegraphics[width=1.0\hsize]{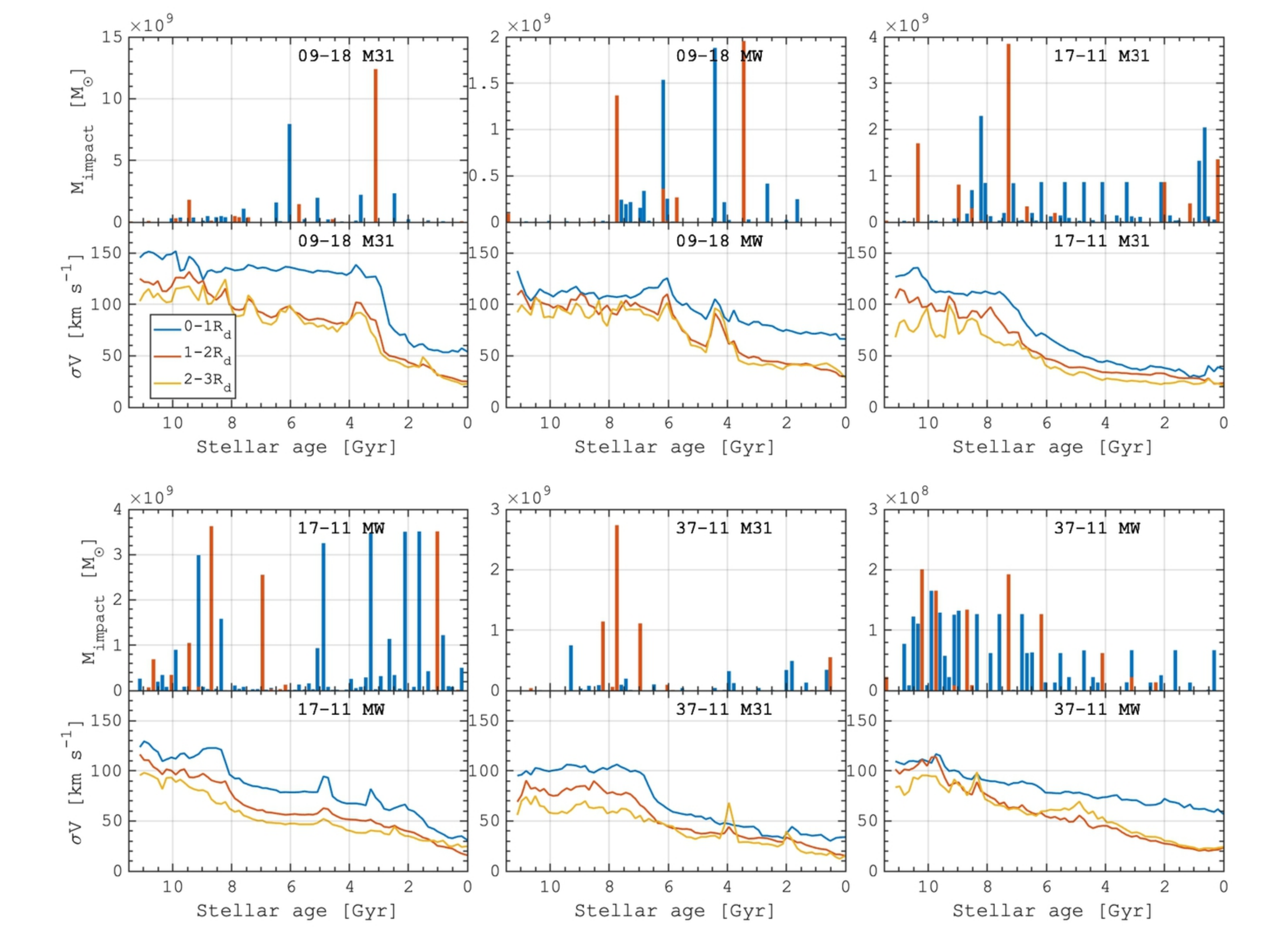}
\caption{Impact of the most significant mergers and close pericentric passages of massive satellites on the age-radial velocity dispersion~($\sigma_R$) relation of M31 and MW HESTIA galaxies. For each galaxy, in the top panels we highlight the time of the close passages~($<10$ kpc from the host) and the stellar mass~($\rm M_{impact}$ at the time of the close passage) of the dwarf galaxies which either merged~(red) or survived at $\rm z=0$~(blue). The bottom panels show the stellar velocity dispersion as a function of the stellar age time for three galactocentric regions: $\rm (0-1)R_d$~(blue), $\rm  (1-2)R_d$~(red), and $\rm  (2-3)R_d$~(green), where $R_d $ is the disc scale length at $\rm z=0$. The growth of the stellar velocity dispersion with the stellar age is remarkable in all the HESTIA galaxies, where in addition to the secular increase likely being caused by the scattering of stars on the spiral arms, we find some age ranges where the velocity dispersion increases faster. These features in stellar kinematics are linked to the impact of the mergers, which heat up the stars that formed before the merger. In the case in which several mergers occurred, the impact of the earliest ones is hard to see because the most recent ones, especially if they are more massive, heat up the disc to the maximal values. }\label{fig1::vel_dispersion_encounters}
\end{center}
\end{figure*}

\begin{figure*}[t!]
\begin{center}
\includegraphics[width=1\hsize]{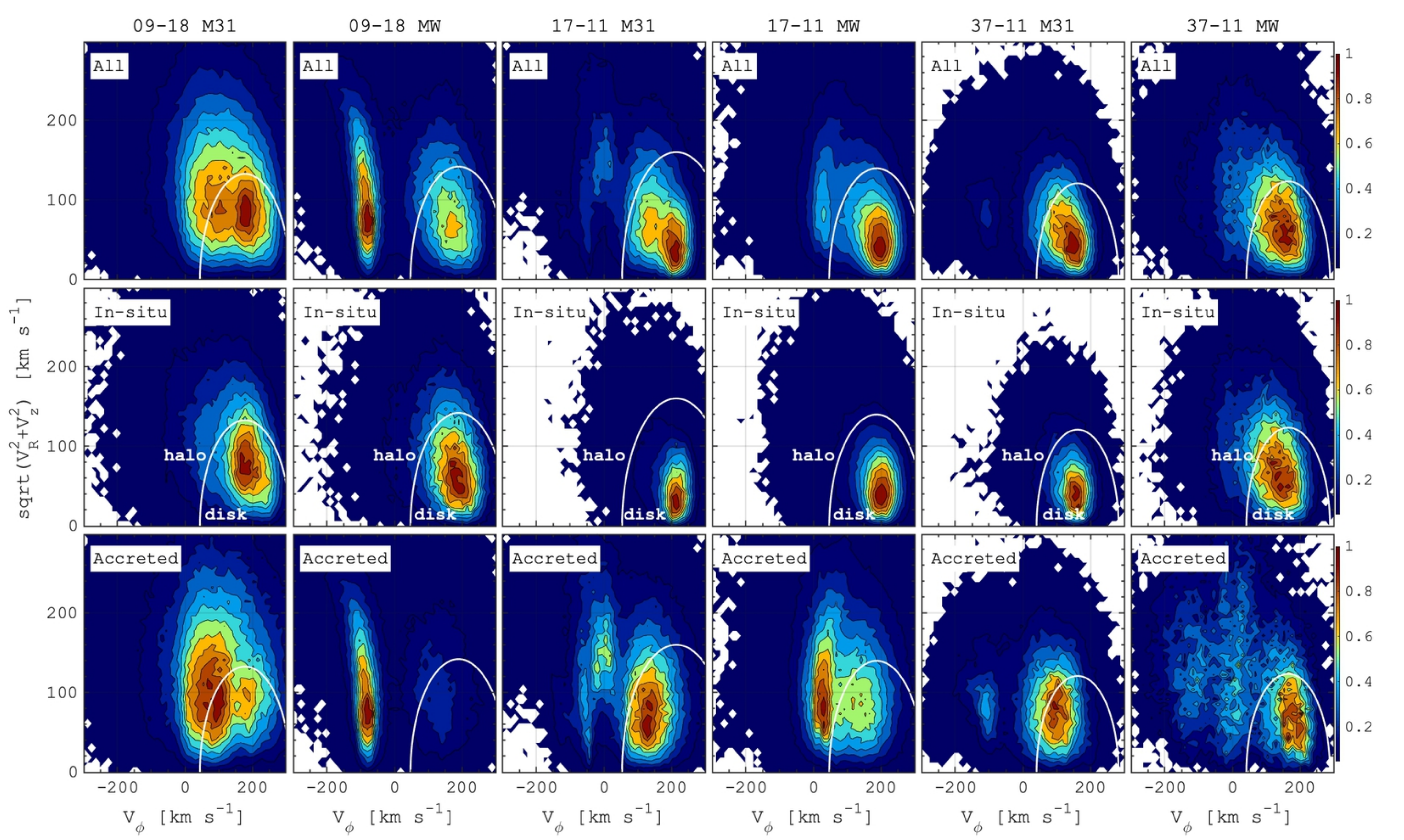}
\caption{Toomre diagram for stars located in $(1-3)R_d$ at $\rm z=0$~($R_d$ is the disc scale length from \cite{2020MNRAS.498.2968L}). The top panels show all of the stars; middle panel corresponds to in situ populations only; and the bottom panel shows all accreted stars. In each panel, the distributions are normalised by the maximum value. White lines correspond to a threshold that is used to distinguish stars with disc-like kinematics from stars with halo-like kinematics~\cite[see, e.g.][]{2010A&A...511L..10N}: $\rm \sqrt{V^2_r+V^2_z+(V_\phi-V_{LSR})^2}<180/240\times V_{LSR}$, where $180$~\kmps is the value typically used in the MW, $240$~\kmps is the LSR in the MW~\citep{2014ApJ...783..130R} and $V_{LSR}$ is the mean rotational velocity in a flat part of the rotation curve in a given HESTIA galaxy at $\rm z=0$. Even taking some overlap into account, the distribution of accreted and in situ stars is very much so different, where in situ stars show mainly disc-like kinematics while the accreted stars show a diverse behaviour: from strongly counter-rotating~(in MW 09-18) to non-rotating~(M31 and MW in 17-11) or weakly co-rotating~(M31 in 09-18), but also mixed kinematical properties~(M31 and MW in 37-11).}\label{fig1::toomre_diagram}
\end{center}
\end{figure*}

Generally speaking, redistribution of the in situ formed stars caused by different types of mergers and close passages of massive systems can also result in an inner stellar halo~\citep{2004ApJ...612..894B,2009ApJ...702.1058Z,2010MNRAS.404.1711P,2011MNRAS.416.2802F,2011MNRAS.415.2652H,2012MNRAS.420.2245M,2013MNRAS.432.3391T, 2015MNRAS.454.3185C, 2015ApJ...799..184P,2016MNRAS.459L..46M}. In the Milky Way~(MW) this issue became topical in light of the recent \Gaia~(ESA) data releases~\citep{2016A&A...595A...2G,2018A&A...616A...1G}. Using the \Gaia~DR1-TGAS sample of local stars, \cite{2017ApJ...845..101B} conclude that locally observed metal-rich ($\FeH>-1$) stars on halo-like orbits have likely been born in situ. Combining the \Gaia DR2 data with chemical abundances from the APOGEE spectroscopic survey \cite{2019A&A...632A...4D} suggest that the vast majority of stars at few kiloparsecs from the Sun with halo-like kinematics have thick disc-like chemical abundances~\citep[see also][]{2018ApJ...863..113H, 2018Natur.563...85H}.

Since the age distribution of the heated population has the same age distribution as the Gaia-Sausage-Enceladus~(GSE) merger~\citep{2019NatAs...3..932G}, it is now accepted that this event played a key role in the formation of the inner stellar halo of the MW~\citep[][]{2018ApJ...863..113H, 2018Natur.563...85H,2020MNRAS.494.3880B}, as well as the disc \citep{lu22, ciuca22, xiang22}. At the same time, the effect of even earlier mergers on the dynamics of the disc is not yet quantified.

A more detailed analysis of the \Gaia DR2 by \cite{2020MNRAS.494.3880B} allowed for the in situ halo~(heated disc) to be defined as the `Splash' stars~\citep[earlier dubbed ``Plume''][]{2019A&A...632A...4D} that have little to no angular momentum with a substantial fraction of stars on retrograde orbits. This was first found by \cite{2012A&A...538A..21S} at the solar vicinity, who analysed the accreted and in situ halo populations from \cite{2010A&A...511L..10N}. This latter work does indeed represent the original discovery of GSE in chemo-kinematic spaces and at a few hundred parsecs from the Sun. According to \cite{2020MNRAS.494.3880B}, the Splash stars can be found around $\FeH<-0.5$ with $V_\phi<80$~\kmps; however, they have a wide range of eccentricities ($0.5-1$, see also \cite{2019MNRAS.482.3426M}), which makes it difficult to use them as a proxy for the merger parameters. It is important to note, that \cite{2020ApJ...891L..30A} have recently proposed a different model of the Splash-like feature formation, where a scattering of massive clumps at high redshift forms a metal-rich, low angular momentum population, without the need for a major merger.

Thanks to a new generation of cosmological hydrodynamic simulations the study of the accretion histories of simulated galaxies is now possible in great detail~\citep[see, e.g.][]{2015MNRAS.446..521S, 2016ApJ...827L..23W,2016MNRAS.457.1931S, 2018MNRAS.473.4077P}; see also the recent review by \cite{2020NatRP...2...42V}. A few previous studies have investigated the effects of the mergers on the host in cosmological hydrodynamical simulations. \cite{2020MNRAS.494.3880B} analysed the Auriga galaxies, demonstrating a sharp change in the host stars' kinematics right after the merger. In particular, the MW-like galaxies show a sharp decrease~(down to zero) in the rotational velocity of stars, which existed in the disc before the massive merger. Using the same set of models, \cite{2020MNRAS.497.1603G} found evidence that gas-rich mergers heated the proto-disc of the simulated galaxies, scattering stars onto less circular orbits, such that their rotational velocity and metallicity positively correlate, thus contributing an additional component that connects the galactic thick disc to the inner stellar halo. \cite{2020MNRAS.497.1603G} also suggest that the counter-rotating fraction of proto-galaxy stars can be used to infer the stellar mass (ratio) of the GSE progenitor. Using a single simulation from EAGLE, \cite{2019ApJ...883L...5B} demonstrate that a GSE-like merger can heat up stars in the early thin disc, forming the present-day thick disc, while the merger debris constitute the stellar halo. Of course, thick discs are much more complicated than what a single event can produce (e.g., the observed strong age gradient in the MW thick disc~\cite{2016ApJ...831..139M}) where the outer parts come from the nested flares of mono-age populations \citep{2015ApJ...804L...9M, 2019MNRAS.487.3946M, 2021MNRAS.501.5105G}.

\cite{2021MNRAS.503.5846R} present a new zoom-in simulation of a MW-like galaxy formation where the last major GSE-like merger deposited stars with high metallicity and non-zero tangential velocities. This component strongly overlaps with the in situ population in a number of parameters, making its identification very uncertain~\citep[see also][]{2017A&A...604A.106J,Pagnini2022}. Using the NIHAO-UHD simulations, \cite{2021MNRAS.500.3750S} found the signature of a low-metallicity in situ population (with both prograde and retrograde rotation) with disc-like kinematics~\citep[see  also][]{2006MNRAS.365..747A,2019MNRAS.484.2166S,2020A&A...636A.115D}. \cite{2021MNRAS.500.3750S} also found that some of the retrograde stars can be formed in situ or have been deposited at very early times. More recently, \cite{2021arXiv210913244D} have shown that about one-third of galaxies from the ARTEMIS cosmological simulations contain GSE-like features which comprise stellar debris from the most massive accreted satellite. \cite{2021arXiv210913244D} also show that the mergers result in a rapid rotation of the disc, seen as the change of the spin orientation -- which tend to be aligned with the merger plane.

The aim of the paper is to investigate the impact of ancient mergers on the in situ disc populations in a set of HESTIA constrained simulations of the Local Group~(LG)~\citep{2020MNRAS.498.2968L}\footnote{https://hestia.aip.de}. In particular, the HESTIA simulations resemble both realistic M31 and MW galaxies in terms of their halo mass, stellar disc mass, morphology separation, relative velocity, rotation curves, bulge-disc morphology, satellite galaxy stellar mass function, satellite radial distribution and the presence of a Magellanic-Cloud-like objects~\citep{2020MNRAS.498.2968L} thus making HESTIA simulations the best tool for studying both environmental effects and the assembly history relevant for the MW and M31 galaxies.

In a series of works based on a new set of HESTIA high-resolution cosmological simulations of the LG galaxies we investigate the phase-space evolution and the present-day structure of the merger debris~\citep[][hereafter \citetalias{KhoperskovHESTIA-2}]{KhoperskovHESTIA-2} and the chemical abundance patterns as a function of stellar ages and kinematics of both accreted and in situ stellar populations~\citep[][hereafter \citetalias{KhoperskovHESTIA-3}]{KhoperskovHESTIA-3}. In this paper, using six M31 and MW analogues from the high-resolution hydrodynamical HESTIA simulations, we focus on how much the mergers contribute to the disc heating, its orbital transformation and the emerging of the Splash/Plume-like stellar populations discovered in the MW. The paper is structured as follows. In Section~\ref{sec1::all_model} we describe the HESTIA cosmological simulations - the initial conditions, physical model, snapshots analysis, and the definition of accreted and in situ stellar populations. In Section~\ref{sec1::assembly_history} we describe some parameters of mergers. In Section~\ref{sec1::results} we analyze the impact of the mergers on the in situ populations focussing on the kinematics and orbital composition of stars. In Sec.~\ref{seq::stellar_halo} we discuss the present-day properties and formation paths of the kinematically defined stellar haloes. Finally, in Section~\ref{sec1::concl} we summarise our main results.

\section{HESTIA simulations}\label{sec1::all_model}

\subsection{HESTIA simulations: Initial conditions}
In this work, we analyse the three highest resolution HESTIA simulations of the LG. Each simulation is tailored to reproduce a number of the LG properties~\citep{2020MNRAS.498.2968L}, including the massive disc galaxies resembling the MW and Andromeda analogues with the population of smaller satellites at $\rm z=0$. Throughout the paper, we analyse six galaxies where realistic present-day properties of the satellite population provide the opportunity to study the impact of the M31- and MW-like accretion history on the in situ formed stellar populations. Next we provide a brief description of the code, physical model, and post-processing of the output. For more details, we refer the reader to HESTIA simulations' introductory paper~\citep{2020MNRAS.498.2968L}.

The HESTIA simulations were performed by using the AREPO code~\citep{2005MNRAS.364.1105S,2016MNRAS.455.1134P}, which solves the ideal magnetohydrodynamics  equations on an unstructured Voronoi mesh with a second order finite volume scheme. Self-gravity and other source terms were coupled to the ideal magento-hydrodynamic equations by operator splitting. Gravitational forces were computed using a hybrid TreePM technique~\citep{2005MNRAS.364.1105S} with two Fourier mesh levels, one for the full box and one centred on the high-resolution region. 

The initial conditions for the run are constrained using  \textit{Cosmicflows-2} (CF2) peculiar velocities \citep{2013AJ....146...86T}. The CF2 catalogue is first grouped \citep{2018MNRAS.476.4362S} and a bias-minimization technique is applied \citep{2015MNRAS.450.2644S}. The technique is based on the Wiener Filter/Constrained Realization \citep{1991ApJ...380L...5H} algorithm, combined with the reverse Zeldovich approximation, spelled out in \cite{2013MNRAS.430..888D,2013MNRAS.430..912D,2013MNRAS.430..902D}.

The HESTIA simulations use the galaxy formation model from \cite{2017MNRAS.467..179G}, which is based on the Illustris model~\citep{2013MNRAS.436.3031V}, and implements the most important physical processes relevant for the formation and evolution of galaxies. It includes cooling of gas via primordial and metal cooling~\citep{2013MNRAS.436.3031V} and a spatially uniform UV background~\citep{2013MNRAS.436.3031V}. The ISM is described by a subgrid model for a two-phase medium in which cold star-forming clouds are embedded in a hot volume-filling medium~\citep{2003MNRAS.339..289S}. Gas that is denser than $\rm n_{thres}=0.13~cm^{-3}$ forms stars following a Schmidt-type star formation law. Star formation itself is done stochastically and creates star particles with the target gas mass that represent single stellar populations. The model includes mass loss and metal return from asymptotic giant branch stars, core-collapse supernovae, and Type Ia supernovae that are distributed in the cells around a star particle. Galactic winds are implemented by creating a wind with a given velocity and mass loading just outside the star-forming phase~\citep{2017MNRAS.467..179G,2019MNRAS.490.4786G,2019MNRAS.490.3234N}. The Auriga model also follows the formation and growth of supermassive black holes and includes their feedback as active galactic nuclei.

The highest resolutions HESTIA simulations are based on the re-runs of the low resolution Dark Matter~(DM) only simulations where two overlapping $3.7$~Mpc ($\rm 2.5\, Mpch^{-1}$) spheres are drawn around the two main LG members at  $z = 0$ and then populated with $8192^3$ effective particles. The mass and spatial resolution achieved is $m_{dm} = 1.5\times 10^5$~\Msun, $m_{gas} = 2.2 \times 10^4$~\Msun\, and $\varepsilon = 220$~pc. HESTIA simulations assume a cosmology consistent with the best fit values~\citep{2014A&A...571A..16P}: $\sigma_8 = 0.83$ and $\rm H0 = 100~h\, km\, s^{-1} Mpc^{-1}$ where $h = 0.677$. We adopt $\Omega_\Lambda = 0.682$ throughout and $\Omega_M = 0.270$ and $\Omega_b = 0.048$. 

\subsection{HESTIA simulations: Post-processing}\label{sec1::actions_calculation}
Halos and sub-haloes are identified at each redshift by using the publicly available AHF\footnote{http://www.popia.ft.uam.es/AHF} halo finder~\citep{2009ApJS..182..608K}. Galaxy and halo histories are estimated via merger trees~(MergerTree tool from the AHF package). At each redshift, accretion events are found by identifying which subhaloes at a given snapshot are identified as 'field' haloes at the previous snapshot. Haloes may also grow via smooth accretion from the environment, namely via gravitationally attracting particles in their vicinity.

For each M31- and MW-simulated galaxy, as well as each snapshot, we define a coordinate system $(x,y,z)$ centred on $10\%$ of the most bound in situ star particles and aligned with the principal axes of this in situ stellar component. The disc plane of the host galaxy thus lies in the $xy$ plane, $R_{gal}$ is the cylindrical galactocentric distance, and the rotation is along the $z$ axis. Our study uses velocity in galactocentric cylindrical coordinates: tangential velocity ($V_{\phi}$), radial velocity ($V_r$), and velocity in the $z$ direction ($V_z$). We also make use of the integrals of motion, focussing on angular momentum in the $z$ direction ($L_z$) and total orbit energy per unit mass $E$. 

To characterise the orbital parameters of stars~(eccentricity, maximum cylindrical distance or apocentres $R_{max}$, and maximal vertical excursion from the disc mid-plane $Z_{max}$), we do not analyse the motion of particles across the snapshots, but integrated the orbits of star particles in a smooth potential. This allowed us to obtain the instantaneous values of the orbital parameters for each snapshot independently, which is not possible from the direct output data. To compute the instantaneous orbital parameters of star particles, first we used AGAMA~\citep{2019MNRAS.482.1525V} to model a smooth gravitational potential. In order to avoid the perturbation of orbits from massive satellites, we interpolated the galaxy potential using only particles associated with the host galaxy. The potential due to dark matter and halo gas is represented by a symmetric expansion in spherical harmonics up to $l = 4$, while the potential of the stars and the gaseous disc was approximated by an azimuthal harmonic expansion up to $m = 4$. We subsequently integrated the orbits of star particles in this potential for $20$ Gyr. This timescale was chosen to account for halo particles with small orbital frequencies. The orbits were integrated with AGAMA, using a Runge-Kutta DOP853 integrator with an adaptive time step.

\section{HESTIA galaxies: Assembly history}\label{sec1::assembly_history}

\subsection{In situ and accreted populations}\label{sec1::in-situ_accreted}
In Fig.~\ref{fig1::density_maps} we show both face-on and edge-on stellar density maps for all the M31 and MW galaxies from the three highest-resolution HESTIA simulations~\citep[09-17, 17-11 and 37-11 from ][]{2020MNRAS.498.2968L}. One can see that all the galaxies exhibit discy stellar components of various sizes surrounded by the smooth, but not featureless, stellar haloes. At the same time, the stellar haloes' morphology~(see bottom panels) reveals a number of tidal streams, shells, and a number of isolated dwarf galaxies, which all together somehow represent the assembly histories of the galaxies. 

The stellar content formation history for the six M31 and MW HESTIA galaxies is presented in Fig.~\ref{fig1::re1_time}, where we show the amount of stellar mass that formed at different distances from the host galaxy over time~\citep[see similar results for the Latte suite of FIRE-2 simulations in][]{2021arXiv211002957C}. One could trace a number of narrow tracks, representing the star formation inside dwarf galaxies that either merged at some point with the host or survived to the present day in the halo. At early times, the assembly histories of HESTIA galaxies are quite complicated due to a number of mergers barely separated in time with the star formation distributed over a large radial range. We note that most stars~($\approx 85\%$ in the MW 37-11 galaxy and $65-68\%$ in other galaxies) are still formed inside the main progenitor, which can be seen as a broad stripe in the bottom of each panel. 

In order to quantify the impact of mergers on the host galaxies, we disentangled `in situ' and `accreted' stellar populations. We consider `in situ' stars as those that were formed gravitationally bound to the main progenitor~(either the MW or M31 analogue), while the rest~(formed bound to other subhaloes) we mark as `accreted'. Throughout the paper, we define a merger event as the accretion of a dwarf galaxy that becomes gravitationally unbound from its own halo and gets bound to the main progenitor, according to the MergerTree tool from AHF. Therefore, all particles associated with a sub-halo right before this event are considered as accreted from a single merger and the last snapshot is used as the time of the merger. According to our definition, the mergers remnants constitute the stellar halo. Meanwhile, all the smaller systems that are still bound at $\rm z=0$, while being inside the virial radius of the main progenitor, represent the population of survived dwarf galaxies of M31 and MW analogues~\citep[same definition as in ][]{2020MNRAS.497.4459F}. In other words, the in situ stars form within the main progenitor branch of the galaxy merger tree, independently of the origin of the star-forming gas. Ex situ stars form outside the main progenitor branch and are subsequently accreted onto the host galaxy through mergers or stripping events.

Since we aim to understand the impact of the GSE-like and even more ancient mergers on the MW-like discs, we focus our analysis on the accretion events, which result in the disruption of dwarf galaxies, while the impact of  satellites orbiting around -- similar to the Sgr dwarf and LMC and SMC systems -- is beyond the scope of the present work. In Fig.~\ref{fig1::sat_orbits} we show the orbital decay of dwarf galaxies (with the stellar mass $>10^6~\Msun$ at $\rm z=0$) as a function of time for all six simulated galaxies. We split the populations according to their fate in the potential of the main progenitor: the left column shows satellites that have been accreted~(mergers) and have become unbound and disrupted, while on the right we present the ones that survived until $\rm z=0$. 

In Fig.~\ref{fig1::sat_orbits}~(left), we can see a number of mergers~($\approx 10-40$) in different galaxies, whose stellar remnants contribute to the diffuse halo component~(see Fig.~\ref{fig1::density_maps}). We note that not all of the tracks in Fig.~\ref{fig1::sat_orbits}~(left) end up in the centre of the main progenitor. This means that the satellites become unbound some time before their core reaches the centre or they are fully disrupted in a shell-like structure without a prominent overdensity close to the galactic centre at the time of the merger. One can see that mergers mainly happen at earlier times because apparently these dwarf galaxies, on average, were closer (or more generally have smaller total energy) to the host at high redshift on average; whereas, while dwarf galaxies, especially at their first pericentric passage close to $\rm z=0$, arrive from larger distances~\citep[see, also, ][]{2018ApJ...863...89S,2021arXiv210911557H, 2022MNRAS.511.2610C, 2022A&A...657A..54B,Dupuy_etal}. 

\subsection{Mergers statistics}\label{sec1::mergers_stats}

In Fig.~\ref{fig1::mergers} we show the stellar mass growth of the main progenitors~(in situ stars, black curves), where the final stellar masses of our six HESTIA galaxies cover the range $\rm (4-8)\times 10^{10}~\Msun$, which is comparable to recent estimates of the MW mass~\citep[see, e.g.][]{2016ARA&A..54..529B}. All the mergers with $\rm M_{stars}>10^6~\Msun$ are marked by the coloured circles. The colour of the circles corresponds to the stellar mass ratio relative to the main progenitor stellar mass~($\rm \mu_{*}$) at the time of merger. The satellites that survived~(bound (sub)haloes, see Fig.~\ref{fig1::sat_orbits}, right) are shown by red lines near lookback time $0$. In this figure we also mark the most significant five mergers as M1-M5 from the earliest to the latest one. We define the most significant mergers according to the relative stellar mass ratio at the time of the merger. The masses of satellite galaxies and stellar merger debris were calculated by using all star particles associated with a given object by the AHF halo finder at the present day or at the time of the infall, respectively. We note that we call mergers significant because not all of them can be classified as `major' mergers (and sometimes not even as massive), especially relative to the present-day mass of galaxies. Nevertheless, using our selection of HESTIA galaxies allowed us to explore the impact of $5\times10^8 - 2\times 10^{10}$~\Msun mergers on the main progenitors' discs. The phase-space-chemistry relations of the merger remnants will be analysed in detail in subsequent works~(\citetalias{KhoperskovHESTIA-2} and \citetalias{KhoperskovHESTIA-3}). 

In Fig.~\ref{fig1::mergers} one can see that the total number of mergers varies in the range of $\approx 10-40$ for different galaxies; however, the number of significant mergers with $\mu_{*}>0.2$~(i.e. at least $1:5$ mergers) is only $1-4$. For most of the galaxies the last significant merger happened $>8$~Gyr ago, similar to our current expectations for the MW~\citep[see][for review and references in the Introduction]{2020ARA&A..58..205H}, with the single exception of the M31 analogue in the 09-18 simulation which, however, is in perfect agreement with the M31 observational data suggesting a major merger ($\approx 2.5\times 10^{10}$ stellar mass) event $2.5-4.5$ Gyr ago~\citep[see, e.g.][]{2018NatAs...2..737D,2019A&A...631A..56B}. On the other hand, there are no significant mergers in the 09-18 MW analogue, where the most massive one has only about $\mu_{*}=0.12$.  Hence we observe a large diversity in the merger history of the simulated M31 and MW analogues in the HESTIA simulations. We remind the reader that the HESTIA simulations of the LG consist of M31 and MW analogues, aimed at reproducing a wide range of M31 and MW parameters and their satellites system, including the presence of LMC- and SMC-like object as well as an M33-like object~\citep{2020MNRAS.498.2968L}.

For the M31 analogues, although the 09-18 simulation reproduces the last major merger very well, in other simulations the last major merger is less massive and happens earlier. This suggests that a significant variance in the satellite/merger mass function and accretion time can be the result of the dwarf merger history and coupling between small- and large-scale seeding in the initial conditions, which is not fully constrained in the HESTIA simulations.

A similar merger diversity is seen across the MW analogues, where the 09-18 and 17-11 simulations show slightly more massive mergers, compared to the estimated masses of different debris (e.g., GSE, Sequoia of $\approx (1-5)\times 10^8~\Msun$~\citep{2020MNRAS.498.2472K,2019MNRAS.490.3426D}). This is likely linked to the longer accretion history of the simulated MW analogues which allows the dwarf galaxy progenitors to evolve longer prior to merging and being completely disrupted. The best candidate, in terms of the mergers parameters, in our sample is the MW analogue from the 37-11 simulation, where most of the mergers happened $>8$~Gyr ago and they have stellar masses~($(1-5)\times 10^8~\Msun$), which is in reasonable agreement with the identified merger debris in the MW. Moreover, this galaxy has a few massive objects at $z=0$, similar to the Sgr, LMC and SMC systems. Therefore, the observed diversity of the merger histories in the HESTIA galaxies makes it possible to test both the evolution of the merger remnants in different environments and the response of the in situ stars to the merger events on different timescales.

\section{Merger impact on the main progenitors}\label{sec1::results}

In this section we seek to explore various manifestations of the significant mergers on the in situ stellar populations. 

\subsection{Spatial axis ratio and $V_\phi/\sigma$ evolution of mono-age in situ populations}
We begin by describing some general properties of the stellar populations of the HESTIA simulations. Various simulations suggest that stars that form earlier, that is to say out of turbulent gas at higher redshifts~\citep[see, e.g.][]{2005A&A...437...69B,2012ApJ...754...48F}, tend to be found with a thicker distribution~\citep[see, e.g.][]{2013A&A...558A...9M, 2013ApJ...773...43B, 2016MNRAS.459..199G}. Therefore, in addition to the upside-down formation of discs, one could expect some extra effects caused by the mergers. Indeed, \cite{2013A&A...558A...9M} show that high-redshift stellar samples are both born hot and are additionally heated by mergers, in their hybrid chemo-dynamical model. Moreover, the authors found that merger-induced large-scale radial migration can redistribute hot-born stars (initially concentrated in the inner parts) and populate a more extended thick disc. 

In Fig.~\ref{fig1::flat_or_round} we show the evolution of the galaxy shape~(the ratio between disc thickness and its radial extension) and the $V_\phi/\sigma_R$ relation for stars in different age bins~(mono-age populations), where $V_\phi$ is the mean rotational velocity and $\sigma_R$ is the radial velocity dispersion of stars in a given age range. The shape of the galaxies is roughly estimated as the mean ratio between the mean vertical and radial positions of star particles of in situ populations. In Fig.~\ref{fig1::flat_or_round} we see that older mono-age populations tend to be thicker and less supported by the regular rotation, however, even the oldest stars show a disc-like behaviour, being flattened towards the galactic plane. However, these older populations do not show substantial rotation compared to the random motions, because the $V_\phi/\sigma_R$ parameter is slightly below unity. With the stellar age decreasing, we see that stellar subsystems become thinner and more rotationally supported~\cite[see, e.g.][]{2014MNRAS.443.2452M, 2016MNRAS.459..199G,2017ApJ...834...27M}.

Fig.~\ref{fig1::flat_or_round} suggests that the mergers moderately affect the axis ratio and kinematic characteristics of the mono-age populations. For most of the merger events we notice only temporal changes in the mono-age populations' shape, which are seen as spikes in the top panels, corresponding to thickening and radial contraction of the mono-age populations. However, soon after the merger, the shape continues to evolve smoothly. The velocity-to-dispersion ratios also show short timescale variations during the mergers. Nevertheless, in the case of the most significant mergers, we detect a substantial change in the kinematics of the mono-age populations selected in the $2$~Gyr range bins. In the M31~(09-18) galaxy, the most massive merger at $\approx 3.5-5$~Gyr~(M5), all intermediate-age mono-age populations~($3-7$~Gyr old) heated up, which is seen in the decrease in the $V_\phi/\sigma$ ratio. The oldest populations are less sensitive to such an impact. A similar behaviour in the massive mergers' impact is seen also in other galaxies, except for MW~(09-18) and  MW~(37-11), which did not experience any massive mergers. 

To summarise, although the HESTIA galaxies we analyse experienced a diverse assembly and merger history, the general trends in the kinematics and spatial distribution of the in situ mono-age populations are essentially the same. In particular, older mono-age populations are thicker and, despite some net rotation, are barely supported by the regular rotation. The overall impact of the mergers seems to be rather modest, but more details about the physical properties of the mono-age populations linked to the impact of the mergers are provided in the subsequent sections. 

\subsection{Merger impact on the star formation history, velocity dispersion and orbits of in situ stars}
\begin{figure*}[t!]
\begin{center}
\includegraphics[width=1\hsize]{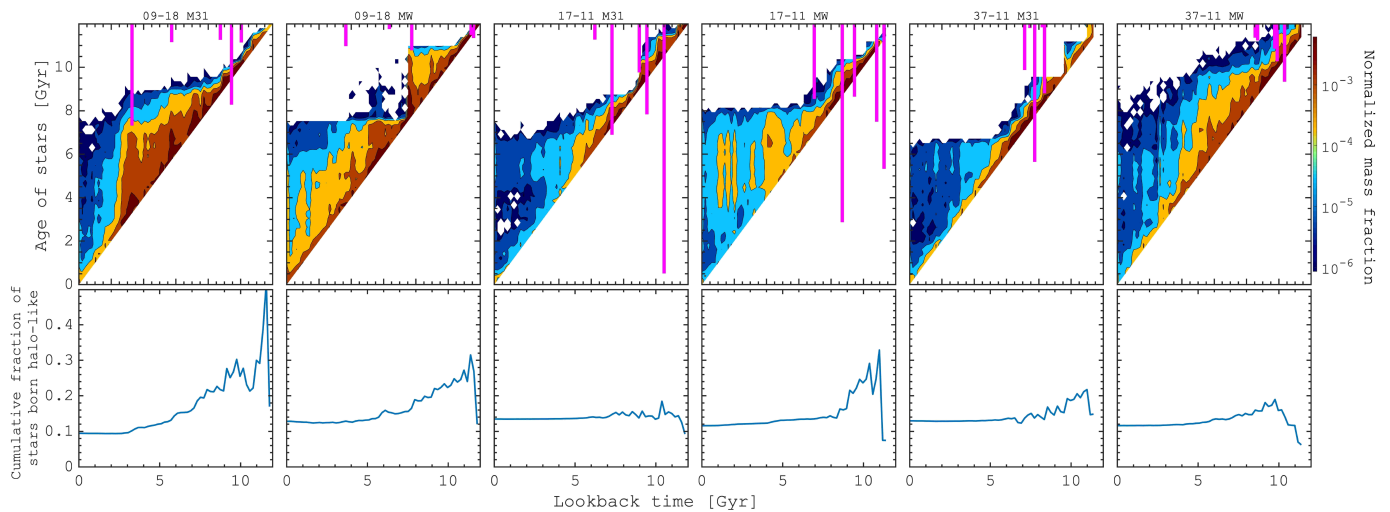}\caption{Formation of the kinematically defined stellar halo. The top panels show the distribution of the ages of stars that start to contribute to the kinematically defined stellar halo~(see Fig.~\ref{fig1::toomre_diagram}) at a given time. Magenta vertical lines highlight five of the most significant mergers~(M1-M5), in terms of the stellar mass ratio. Bottom panels show the cumulative fraction of stars that formed with halo-like kinematics. The figure suggests that at a given time, the newly formed stars contribute the most to the kinematically defined stellar halo region of the Toomre diagram. The reason for this type behaviour is that the youngest stars have colder kinematics compared to the pre-existing stars, thus, these recently that formed stars are the most sensitive to the external perturbations of the disc caused by the mergers. Since the fraction of stars formed kinematically hot does not exceed $10-12\%$ of the in situ stars, the heating of the pre-existing populations is the dominant mechanism of the in situ halo formation in the HESTIA galaxies.}\label{fig1::origin_of_halo_from_toomre}
\end{center}
\end{figure*}

\begin{figure}[t!]
\begin{center}
\includegraphics[width=1\hsize]{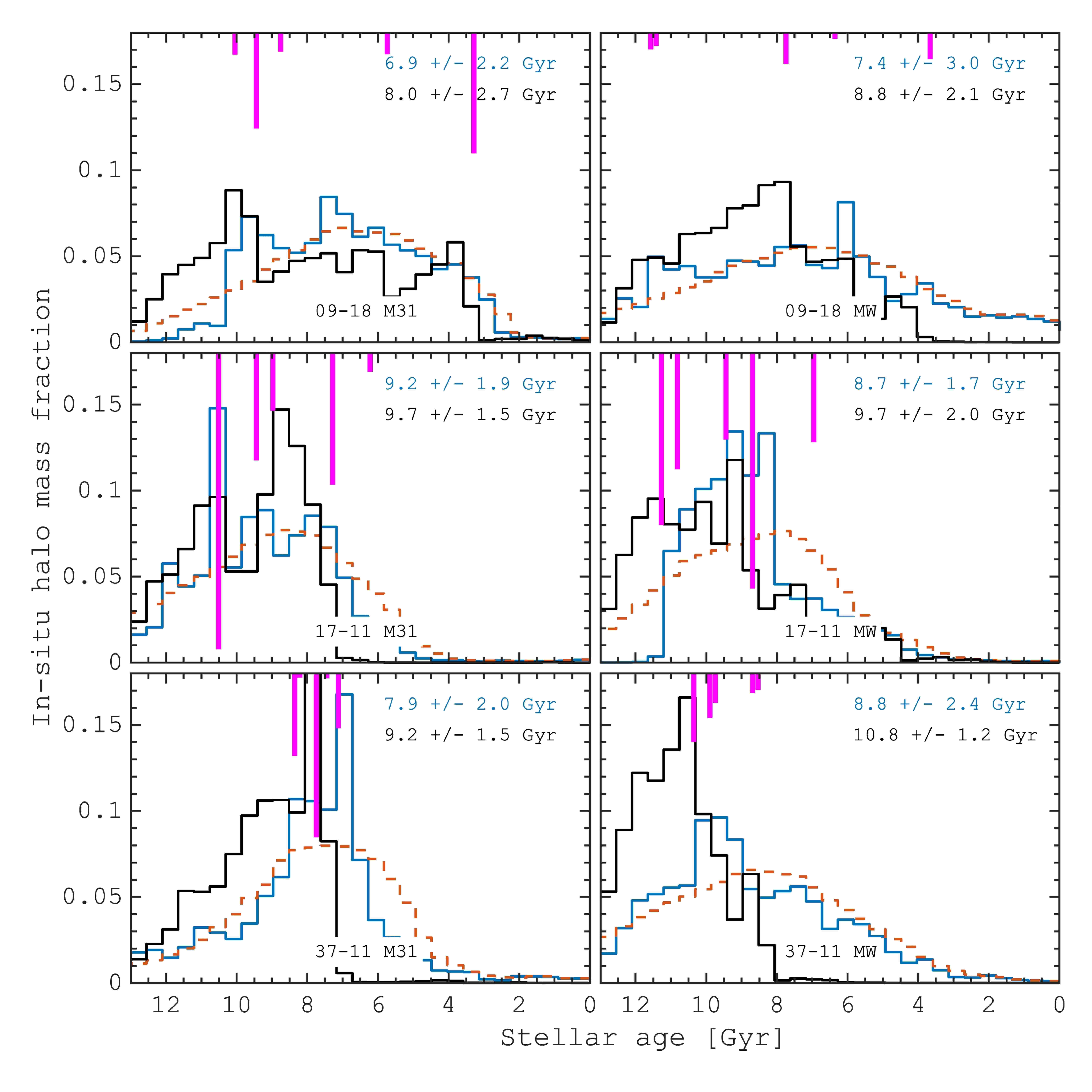}\caption{ Age distribution of the kinematically defined stellar haloes~(blue lines; see, Toomre diagram in Fig.~\ref{fig1::toomre_diagram} for the definition). The red dashed lines correspond to the distributions with the constant $20\%$ age uncertainty. Black lines show the age distribution of the accreted component of the stellar halo. The numbers in the top left corner of each panel show the mean and the standard deviation of the stellar ages for the in situ~(blue) and accreted~(black) halo. The vertical magenta lines highlight five of the most significant mergers~(M1-M5), in terms of the stellar mass ratio. The age distributions show a larger fraction of stars that formed inside the main progenitors close to the times of the mergers. However, the features completely vanished once the age errors were taken into account. Such a distribution transformation is not the result of the error itself but also because a substantial fraction of stars populating the kinematically defined halo between the merger events. }\label{fig1::ages_of_halo_from_toomre}
\end{center}
\end{figure}

\begin{figure*}[t!]
\begin{center}
\includegraphics[width=0.49\hsize]{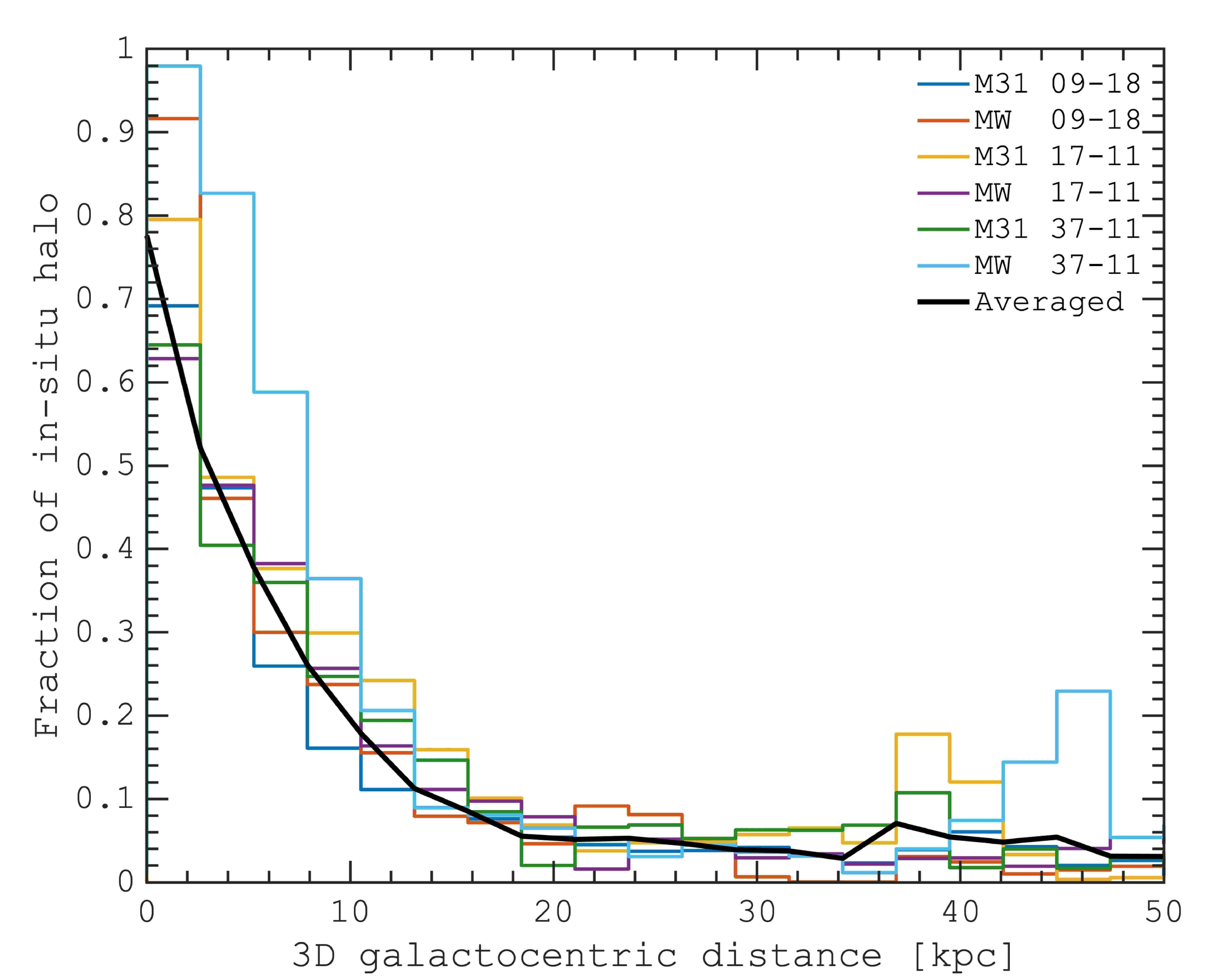}
\includegraphics[width=0.49\hsize]{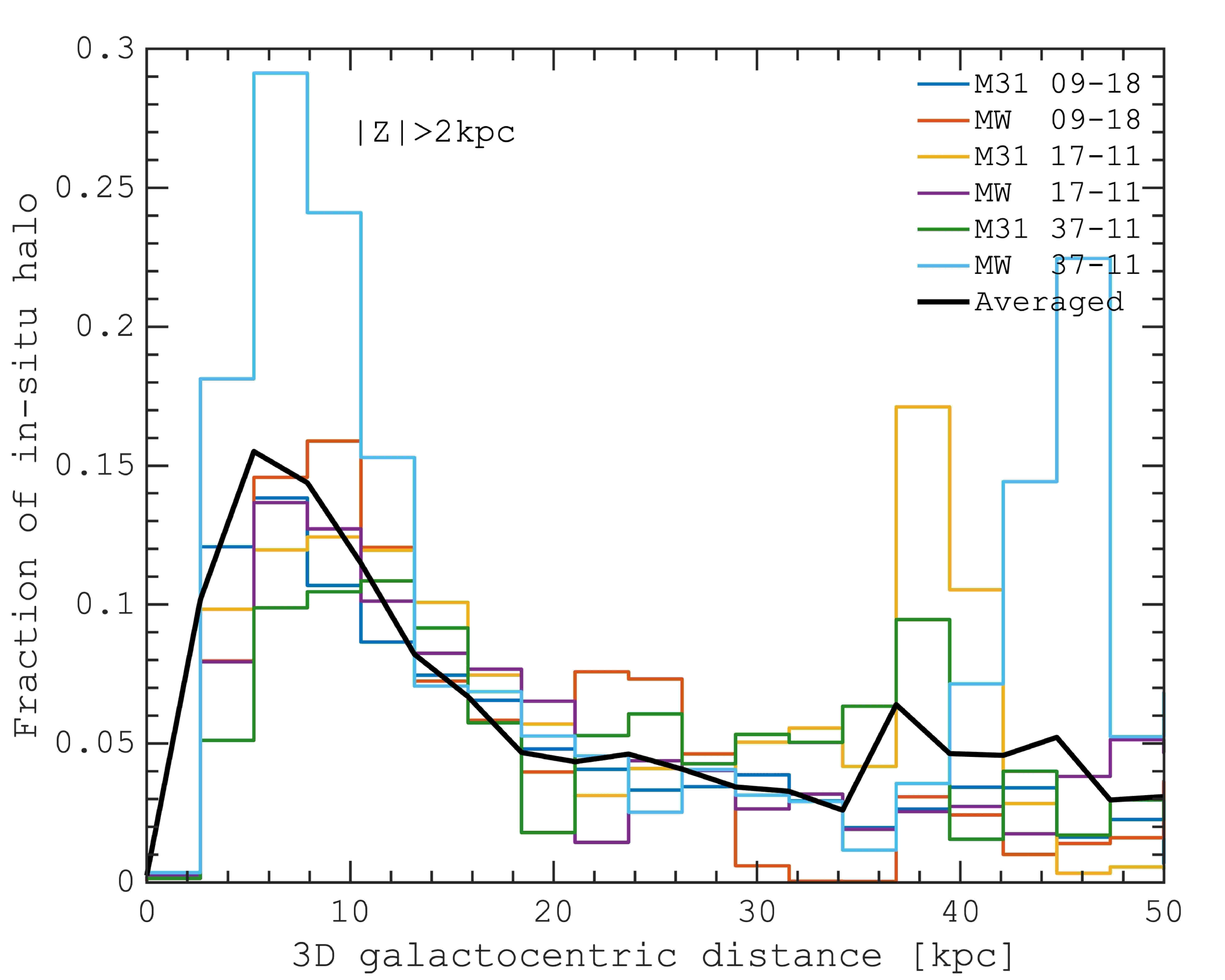}\caption{Relative fraction of the in situ stellar halo as a function of the 3D galactocentric distance in different models at $\rm z=0$. The left panel shows the data without any spatial selections, while in the right panel we adopted $|z|>2$~kpc selection. The in situ halo is defined above the white line in the Toomre diagram in Fig.~\ref{fig1::toomre_diagram}. In all the HESTIA galaxies, similar to the results of the H3 survey of the MW halo~\citep{2020ApJ...901...48N}, the in situ halo contribution rapidly decreases beyond $\approx 10$~kpc and does not exceed $2-5\%$ at larger distances. We note that recent massive mergers in M31~(09-18) and M31~(17-11) galaxies~(see Fig.~\ref{fig1::mergers}) result in the overdensity of in situ populations at $\approx45$~kpc.
}\label{fig1::in_situ_halo_fraction}
\end{center}
\end{figure*}

\begin{figure*}[t!]
\begin{center}
\includegraphics[width=1\hsize]{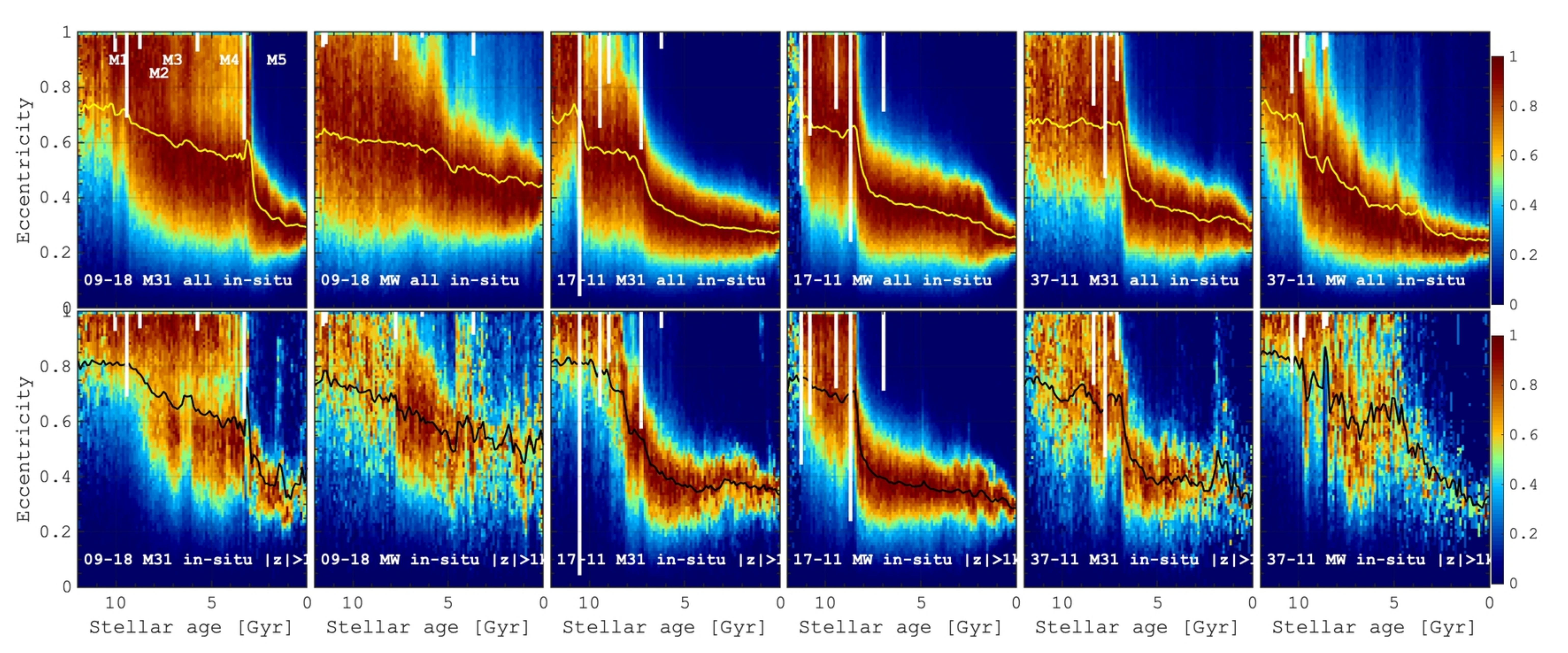}\caption{Variations of the orbital eccentricity distributions as a function of the stellar age in M31 and MW HESTIA galaxies. Distributions were normalised by the maximum value at a given age. The top row corresponds to all in situ stars while the bottom corresponds to in situ stars $>1$~kpc away from the galactic plane mostly excluding the disc component. White vertical lines highlight five of the most significant mergers~(M1-M5), in terms of the stellar mass ratio. The mean eccentricity values are shown by the yellow~(top) and black~(bottom) lines. Similar to the rotational velocity behaviour~(see Fig.~\ref{fig1::in_situ_lz_evolution}), the distributions show several episodes of a sharp increase in the eccentricity which coincide with the mergers. }\label{fig1::in_situ_ecc_evolution}
\end{center}
\end{figure*}

\begin{figure*}[t!]
\begin{center}
\includegraphics[width=1\hsize]{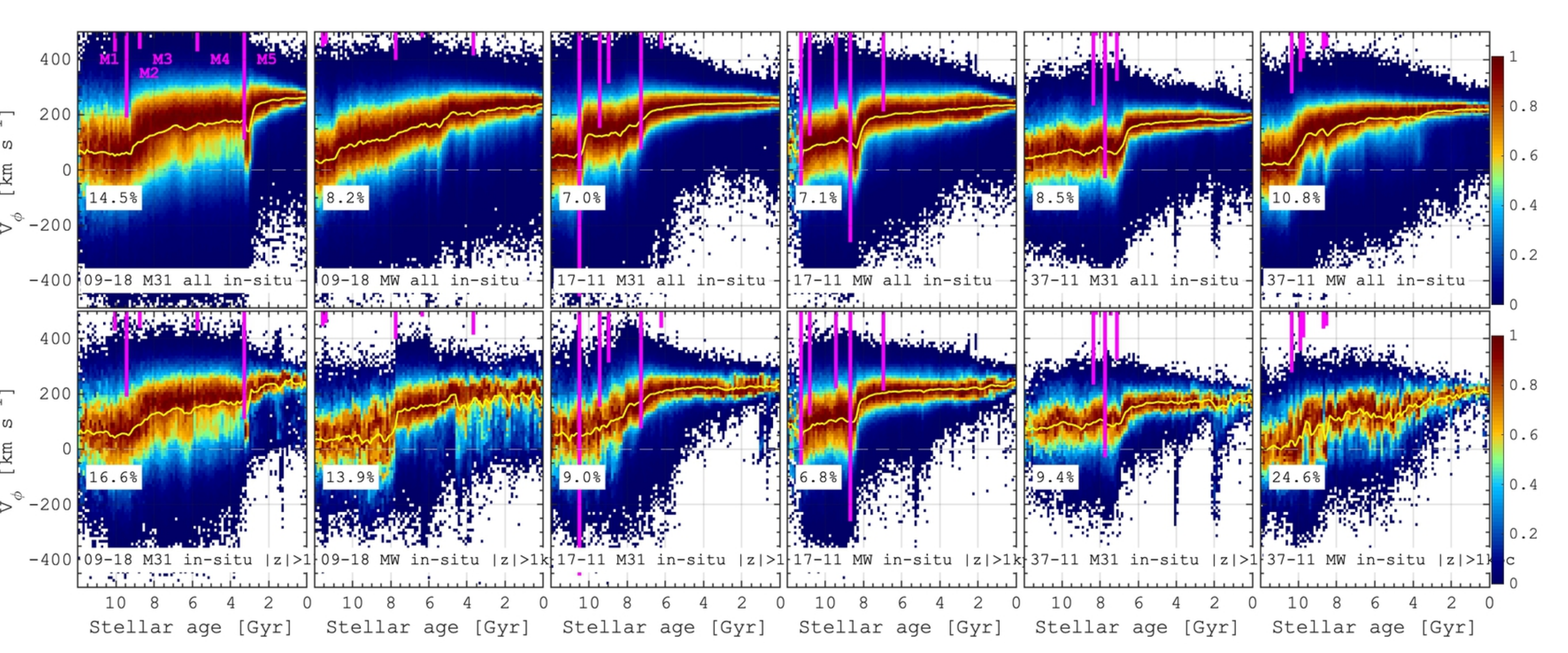}\caption{Structure of the azimuthal velocity as a function of the stellar age in M31 and MW HESTIA galaxies. Distributions were normalized by the maximum value at a given age. The top row shows all in situ stars, while the bottom one corresponds to the in situ stars $>1$~kpc away from the galactic plane, mostly excluding the disc component. Magenta vertical lines highlight five of the most significant mergers~(M1-M5), in terms of the stellar mass ratio, where the length of the line shows the stellar mass ratio of the merger where the vertical panel's size is assumed to be unity. The mean $V_\phi$ values are shown by the yellow lines. The numbers in white boxes show the mass fraction of counter-rotating stars~(negative azimuthal velocity) among the sample presented in a given panel. The azimuthal velocity generally decreases with the stellar age due to secular disc heating; however, at the time of the mergers, it falls faster, resulting in slowly rotating stellar populations similar to the Splash/Plume stars discovered in the MW. We note, however, that in all of the galaxies, except for M31 and MW in the 37-11 simulation, the impact of several mergers is clearly seen in the stair-step-like distributions. }\label{fig1::in_situ_lz_evolution}
\end{center}
\end{figure*}

\begin{figure*}[t!]
\begin{center}
\includegraphics[width=1\hsize]{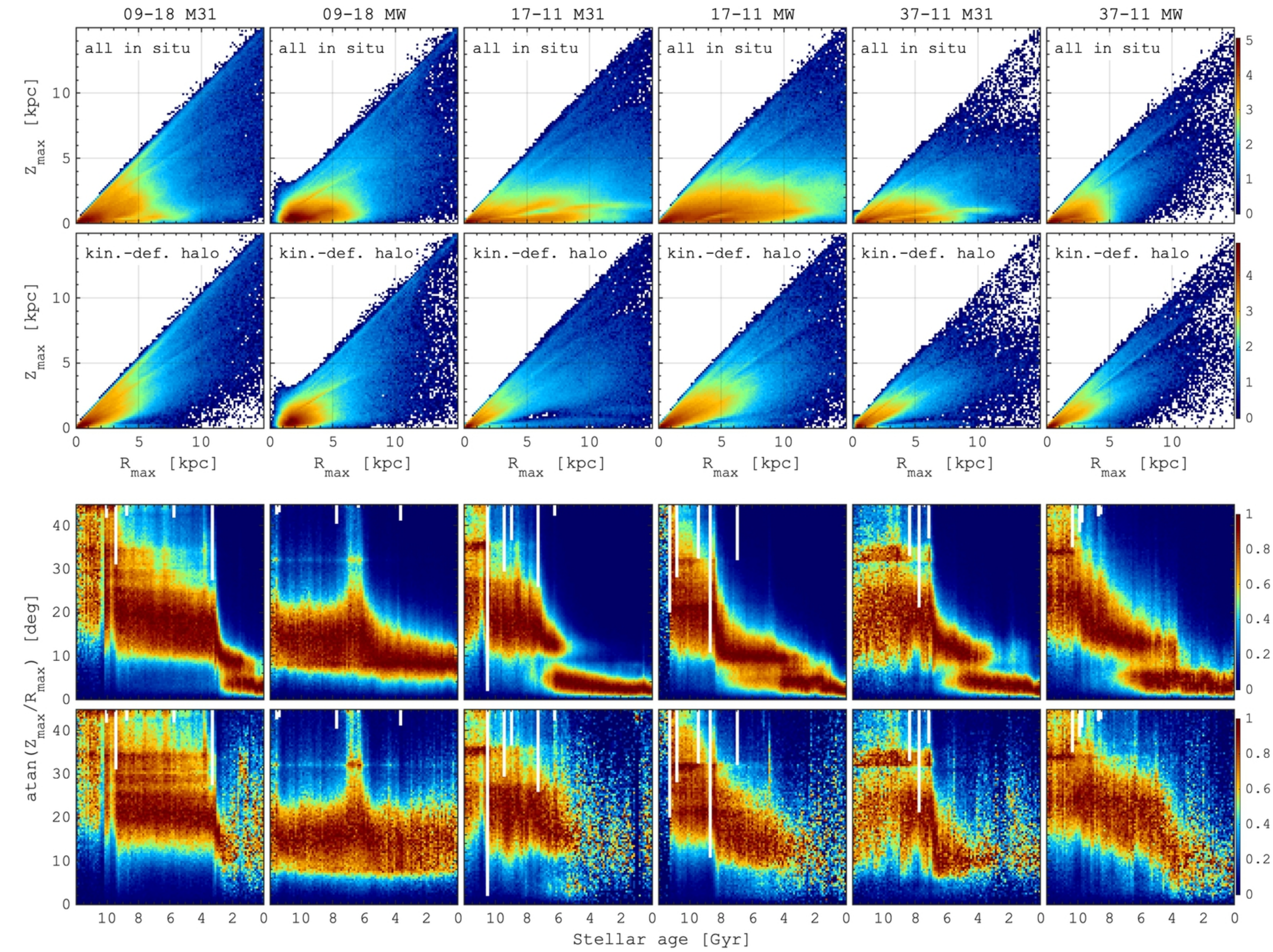}\caption{Ortbital structure of in situ stellar halo. {\it Top:} Stellar density distribution in $\rm R_{max}-Z_{max}$ coordinates for all in situ stars~(first row) and kinematically defined~(see Toomre diagram in Fig.~\ref{fig1::toomre_diagram}) in situ stellar halo~(second row). {\it Bottom:} Distribution of the angles between $\rm R_{max}-Z_{max}$ as a function of the stellar age for all in situ stars~(third row) and kinematically defined stellar halo~(fourth row). The white vertical lines highlight five of the most significant mergers~(M1-M5), in terms of the stellar mass ratio. Similar to a number of recent MW studies~\citep[see, e.g,][]{2018ApJ...863..113H,2021A&A...647A..37K} the HESTIA galaxies reveal a number of wedges in $\rm R_{max}-Z_{max}$ coordinates. The bottom rows show how these wedges emerge over time. We find that the individual structures in $\rm R_{max}-Z_{max}$ are made of stars predominantly made of stars that formed in between the significant merger events, suggesting a similar origin for similar structures in the MW.}\label{fig1::rmax_zmax}
\end{center}
\end{figure*}

Pericentric passages of satellites may enhance the star formation either via contraction of the ISM or delivering more gas and thus fueling the star formation in the host~\citep{1994ApJ...425L..13M,1995ApJ...448...41H}. Although some observations of the MW rely on this idea~\citep[see, e.g.][]{2020NatAs...4..965R}, some simulations suggest that mergers and galaxy interactions do not always trigger starbursts in the hosts~\citep{2007A&A...468...61D,2008A&A...492...31D,2008MNRAS.385L..38M,2017MNRAS.465.1934F,2021MNRAS.506..531D,2022arXiv221017054A}. To test this issue in the HESTIA galaxies, in Fig.~\ref{fig1::sfr_encounters} we show the star formation histories~(SFHs) of the M31 and MW analogues where we plot the age distribution weighted by the initial mass of star particles. Therefore, the mass loss due to stellar evolution does not affect the SFHs. We split the SFHs into three groups for the stars that are located in different radial bins $(0-1) R_d$~(blue), $(1-2) R_d$~(red), and $(2-3) R_d$~(yellow) at $\rm z=0$, where $R_d$ is the stellar disc scale length~\citep{2020MNRAS.498.2968L}. For each SFH, in the upper subpanels, the vertical lines show the time of close passages for dwarf galaxies (blue) and the mergers~(red), where the height of the lines shows the total stellar mass of a perturber at that time~($\rm M_{impact}$).  In this figure, close passages include both pericentric passages and fly-bys of massive systems if they are closer than $10$~kpc from the centre of the host galaxy.

One can see a number of peaks in the SFHs where some of them apparently correlate with the mergers or close encounters of dwarf galaxies. In most cases, we see a clear correlation between the mergers or close encounters and the peaks of the star formation in the hosts. However, the amplitude of the bursts does not seem to correlate with the mass of the perturbed, and even more extreme in many cases, passage of a massive system that does not always trigger star formation in the host.  This behaviour is somewhat predicted by some other models depending on the interaction parameters and the amount of gas available for star formation~\citep[e.g.,][]{2007A&A...468...61D,2008A&A...492...31D,2017MNRAS.465.1934F}. Another issue we can address is that most star formation bursts are seen for the stars residing in the inner parts of the galaxies~(blue lines); whereas, the outer discs are likely less affected where, however,  some bursts can be seen but have lower amplitudes. Overall, we suggest that if the peaks of the star formation are found in the host galaxies, they correlate with close passages of dwarf systems of the mergers. A more detailed analysis of the external impact on the star formation activity in the HESTIA galaxies, including the star formation inside the dwarf galaxies, is a subject of other work~\citep[see, e.g.][]{2020MNRAS.497.1603G,2022MNRAS.513.1867D}.

Next, we address the impact of the mergers on the orbital composition of in situ stellar populations.  In Fig.~\ref{fig1::rmax_zmax_encounters} we show the mean $\rm Z_{max}$ and mean $\rm  Z_{max}/R_{max}$ ratio as a function of stellar age. Overall, we see a gradual increase in both $Z_{max}$ and $Z_{max}/R_{max}$ with stellar age. At the same time, we can see that merger times coincide with a sharp increase in the mean $\rm Z_{max}$ for all of the stars that existed in the host galaxy prior to the merger. Although the latest mergers seem to cause the largest increase in mean $Z_{max}$, in some galaxies one can still detect a few peaks of the vertical excursion of stars correlating with previous mergers. Interestingly, mergers also affect the $\rm Z_{max}/R_{max}$ ratio, which is typically $0.1-0.2$ for the youngest populations. However, for the stars that experienced the merger impact, this ratio can reach up to $0.5-0.6$. This result suggests that the mergers are relatively more efficient at disc heating in the vertical direction, which inevitably causes disc flaring \citep{2015ApJ...804L...9M}.

Another issue is the impact of the mergers on the stellar velocity dispersion. As mentioned in the Introduction, it is expected that the GSE merger in the MW heated the pre-existing disc, resulting in the Plume/Splash feature~\cite[see][]{2019A&A...632A...4D,2020MNRAS.494.3880B}. In Fig.~\ref{fig1::vel_dispersion_encounters} we show the age-radial velocity dispersion relation, ($\sigma_R$), for stars in different radial bins, as in the previous figures. In addition to the monotonically decreasing velocity dispersion with diminishing age, we can identify two different signals. Firstly, a number of peaks correlate with the times of external perturbations, suggesting that some mergers mostly affect  stars that formed during or just before the interaction. Alternatively, these stars could have been formed in the tidal tails caused by the perturbations and thus have a locally higher velocity dispersion. The second type of feature is a step in the velocity dispersion, which results in a flattening for all stars that existed in the disc before the merger took place. This appears to be the result of the most significant mergers. Such a step in the MW age-velocity relation has long been associated with the effect of the last massive merger \citep{quillen01, freeman02, minchev18}, now known as the GSE event.

\section{The emergence of the in situ stellar halo}\label{seq::stellar_halo}

\subsection{Kinematically defined stellar haloes}

In this section, from the overall impact of the mergers on the discs, we move towards the analysis of the in situ stellar halo. Different definitions of the stellar halo have been adopted in the literature and here we use the one purely based on the stellar kinematics~\citep{2010A&A...511L..10N}. In particular, we analyse the structure of the Toomre diagram. In Fig.~\ref{fig1::toomre_diagram} we show the Toomre diagram for all stars~(top), in situ only~(middle), and all stars from the merger remnants~(bottom, accreted only) at $\rm z=0$. To enhance the separation between the disc and the halo populations, we have omitted the centres of the HESTIA galaxies and consider the stars that are located in $(0.5-3)~R_d$, which is accessible for the MW galactic surveys. In all the galaxies in our sample, the distributions on top show the presence of two distinct components, where one is centred near the Local Standard of Rest (LSR) and the others tend to show a slow prograde or even retrograde rotation with larger random velocities. The latter ones are mainly made of stars from disrupted dwarf galaxies, presented in the bottom panel which we explore in detail in \citetalias{KhoperskovHESTIA-2}. Therefore, the subsequent the focus of our study is to explore the properties of the in situ stars, which, being weighted towards the LSR, show significant extension~(see middle panels).

First, we look at the formation path of the in situ stellar halo, defined kinematically in the Toomre diagram. In particular, we investigate when the stars appear outside of the white circle in Fig.~\ref{fig1::toomre_diagram}, which is often used as a threshold between the disc and halo stars in the MW. In Fig.~\ref{fig1::origin_of_halo_from_toomre}, we present the origin of the stellar halo where the $\rm X$ axis corresponds to the time when certain in situ formed stars appear above the white line in Fig.~\ref{fig1::toomre_diagram}, and the $\rm Y$ axis shows the age distribution of these stars. First, we can see that the kinematically defined in situ stellar halo continues to grow over time in all of the HESTA galaxies. This is likely due to constant perturbations from both orbiting dwarf galaxies and ancient mergers. This result should be taken into account in the MW analysis where the GSE-like merger is believed to kick stars out of the disc, and all of the halo stars~(both in situ and accreted) that we observe gained  halo-like motions at the time of the merger. In HESTIA galaxies, substantial halo stars arrive at the kinematically defined halo region some billion years ago after the mergers. This suggests that the overall mass of the in situ halo cannot be used to constrain the impact of the mergers directly. Nevertheless, in most cases, the massive mergers impact the in situ populations, making them move similarly to halo stars. This is seen as the enhancement of the density~(fraction of the overall kinematically defined stellar halo) in Fig.~\ref{fig1::origin_of_halo_from_toomre} close to the mergers. The bottom panels of Fig. \ref{fig1::origin_of_halo_from_toomre} show the cumulative fraction of the in situ stars, which have halo-like kinematics at the time of their formation. This fraction can be rather high~(0.2-0.5) at the early phases of the galaxy assembly. However, it decreases rapidly over time, and at the present day, only about $10\%$ of the in situ stars represent a genuine stellar halo. The rest is the result of the heating of the disc stars, thus suggesting that the heating of the pre-existing populations is the major source of the in situ stellar halo formation.

Another interesting feature we can notice in the Fig.~\ref{fig1::origin_of_halo_from_toomre} is that the youngest stars are the most sensitive populations to the impact of the mergers. At a given time, more of the contribution to the halo is from the stars that formed recently, which is seen as the overdensity along the one-to-one line. This is not expected for the gradual heating of the disc where stars cannot jump from the cold orbits to the halo-like ones without passing through the phase of moderate heating. An alternative opportunity is the external impact~(tidal interaction with satellites and mergers), which affects populations on colder orbits more efficiently. However, the most massive mergers still affect even warm-orbit stars that formed several Gyr before but already experienced some heating. This is seen best in the  case of the M5 merger at $\approx 3$~Gyr in the M31~(09-18) galaxy, but this is also noticeable in other models.

In order to test whether it is possible to date the mergers by using the in situ stellar halo formation history in Fig.~\ref{fig1::ages_of_halo_from_toomre}~(blue lines), we show the age distribution of the kinematically defined in situ stellar halo~(see Fig.~\ref{fig1::toomre_diagram}) while the black lines show the accreted component. Interestingly, our definition of the in situ stellar halo results in a wide range of stellar ages where some narrow peaks are clearly associated with the mergers~(vertical magenta lines). The peaks in the distribution of the stellar ages are not very prominent because, once we assumed a constant $20\%$ age uncertainty~(red dashed lines), they were completely erased, which is the result of a continuous inflow of stars to the kinematically-defined region in the Toomre diagram. On average, the in situ stellar halo of the HESTIA galaxies is rather young with the mean stellar age of $8-9$~Gyr (a single exception is M31 in the 09-18 simulation with a recent massive merger) which is $\approx 1$~Gyr younger compared to the accreted component. 

Finally, we measured a spatial distribution of the in situ stellar halo in such a way it could be compared to the MW data. In Fig.~\ref{fig1::in_situ_halo_fraction} we show the relative fraction of the in situ stellar halo, defined kinematically using the Toomre diagram in Fig.~\ref{fig1::toomre_diagram}. In the left panel, we present the in situ halo mass fraction without spatial cuts, while on the right we show only the numbers avoiding the disc region~($|z|>2$~kpc). The distributions show that the in situ~(heated) populations represent $\approx5-30\%$ in the imminent vicinity of the disc; this value decreases drastically at distances larger than $10-15$ from the galactic centre and stays almost constant~($2-5\%$) out to $50$~kpc. The rest of the stellar halo is made of merger debris. The right panel can be directly compared to the results delivered by the H3 survey~\citep{2020ApJ...901...48N}, where the MW in situ halo fraction decreases from $\approx 15\%$ at $2-5$~kpc to $\approx 5\%$ at $15-20$~kpc. This suggests that, despite the relatively more massive accreted systems, the HESTIA galaxies represent the MW-disc stellar halo and disc behaviour well, and thus the conclusions we make about the origin of the in situ stellar halo can be further explored in the MW.

\subsection{Searching for Splash/Plume features in HESTIA galaxies}
Here, we link our finding to the known impact of the GSE merger which is believed to be the major heating event in the MW. First, in Fig.~\ref{fig1::in_situ_ecc_evolution} we present the orbital eccentricity distribution variation as a function of stellar age for all in situ stars~(top) and for in situ stars $>1$~kpc away from the disc plane~(bottom). Similar to the previous figures, the most significant mergers are marked by the vertical lines where the length of the lines corresponds to the merger stellar mass ratio~($\mu_*$) where the panel size is assumed to be unity. The yellow lines show the mean eccentricity trends. Fig.~\ref{fig1::in_situ_ecc_evolution} clearly shows that the most massive mergers~(the longest white vertical lines) result in a sharp increase in the mean eccentricity of the stars. In all of the HESTIA galaxies, the in situ formed halo stars show the breaks of the eccentricity distribution near the significant mergers. More generally, the most massive mergers lead to the substantial heating of the in-plane orbits of stars that can be found on the radial orbits with an eccentricity above $0.8$. We note also, that the in situ halo stars have a broad distribution of eccentricities and even some of the oldest stars can be found on nearly circular orbits~\citep[see, e.g.][]{2019MNRAS.484.2166S}.

\cite{2010A&A...511L..10N} showed that the thick disc stars with halo-like kinematics found in solar vicinity data could be the result of the heating of the disc. More recently, using Gaia data \cite{2019A&A...632A...4D} and \cite{2020MNRAS.494.3880B} quantified the impact of the GSE merger on the pre-existing MW stellar populations that are rotating more slowly, and even counter-rotating relative to disc stars. The excess of the slowly and counter-rotating (chemically defined) in situ stars is called Splash~\citep{2020MNRAS.494.3880B} or Plume~\citep{2019A&A...632A...4D}. In order to find similar features in the HESTIA galaxies in Fig.~\ref{fig1::in_situ_lz_evolution} we show the variations of the azimuthal velocity as a function of stellar age~(the dependence on the metallicity is discussed in \citetalias{KhoperskovHESTIA-3}) for all in situ stars~(top) and in situ stars $\rm >1kpc$ away from the disc plane~(bottom). The picture here is somewhat similar to the eccentricity distribution in Fig.~\ref{fig1::in_situ_ecc_evolution} where the merger events heat up the orbits of stars, thus decreasing their net orbital motion, and even pushing some of them on counter-rotating orbits. In Fig.~\ref{fig1::in_situ_lz_evolution}, one can see that the HESTIA galaxies reveal a few episodes of disc heating, resulting in the formation of Splash/Plume-like stellar populations. For instance, in the 17-11~M31 galaxy a one-to-one M1 merger $\approx 10.5$~Gyr ago results in the populations with $V_\phi\approx 50$~\kmps net rotation. The second massive merger at $\approx 7$~Gyr ago~(M5) creates the stars with $V_\phi\approx 100$~\kmps net rotation. It is important to note, however, that a very massive merger is not required for the formation of Splash/Plume-like populations. In particular, even a $9\%$ stellar mass merger~(M3) at $\approx 8$~Gyr ago in the 09-18~MW galaxy is able to create a sharp decrease in the azimuthal velocity which is better seen for the halo stars~(bottom row). Among the heated populations, we also notice the presence of stars with negative rotational velocity. Although the counter-rotating stars constitute between $7\%$ to $24\%$ of the stellar halo~(see numbers in the bottom row in Fig.~\ref{fig1::in_situ_lz_evolution}), these populations represent neither classical counter-rotating discs nor any distinct components of the galaxies while being the tails of the heated Splash/Plume-like stars.  

Another interesting feature common in all of the HESTIA galaxies is that even strongly heated stellar populations~(low $V_\phi$, or high eccentricity) still demonstrate a substantial net rotation with the direction aligned with most of the disc stars. This means that even the oldest stars inherit disc-like kinematics, suggesting a very fast collapse of the main progenitor at early times of the evolution and thus leaving no space for a non-rotating in situ stellar halo. Therefore, the stellar halo in HESTIA galaxies is predominantly made of disc stars heated by a few significant mergers.

\subsection{$R_{max}-Z_{max}$ composition of the in situ stellar halo}

In this section, we analyse the orbital composition of the stellar component of the HESTIA galaxies. 
For all star particles, we have reconstructed their orbital parameters by integrating their orbits in the fixed galactic potential at $z=0$ using the AGAMA code~\citep[][see Sections.~\ref{sec1::actions_calculation} for more details]{2019MNRAS.482.1525V}. In Fig.~\ref{fig1::rmax_zmax} we show the density distributions in the $R_{max}-Z_{max}$ plane, where $R_{max}$ is the apocentre and $Z_{max}$ is the maximum height from the mid-plane. Similar to a number of  MW studies we present the distribution of stellar density in $\rm  R_{max}-Z_{max}$ coordinates~(top)~\citep[see, e.g,][]{2018ApJ...863..113H,2021A&A...647A..37K}. In the top panels, we detected a few diagonal overdensities~(or `wedges') separated by the prominent gaps. \cite{2021A&A...647A..37K} show that in the MW, these structures could be due to resonances and the result of chaotic diffusion while \cite{2018ApJ...863..113H} suggest that the discrete wedges in the $\rm  R_{max}-Z_{max}$ plane – are all reminiscent of some impulsive heating of the early Galactic disc related to some accretion event(s).  To clarify the composition of the $\rm  R_{max}-Z_{max}$ plane in the HESTIA galaxies, in Fig.~\ref{fig1::rmax_zmax}~(bottom) we show the distribution of angles $\rm \arctan{Z_{max}/R_{max}}$ as a function of stellar age. In most of the HESTIA galaxies, the lowest-angle wedges appear after the latest significant merger. Some other wedges~(horizontally aligned overdensities in the bottom panels) are made of stars that formed between some other, previous significant mergers. Therefore, we suggest that in the HESTIA galaxies, we detected a correlation between mergers and wedges in the $\rm  R_{max}-Z_{max}$ plane. However, some overlap between these structures and the uncertainties of stellar ages may complicate matching the wedges with accretion events in the MW. 

\section{Summary}\label{sec1::concl}

We have analysed six M31 and MW analogues from the HESTIA suite of cosmological hydrodynamics zoom-in simulations tailored to reproduce the properties of the LG galaxies. In this work, we have focussed our analysis on the impact of the mergers on the shape, kinematics and orbital composition of the in situ populations and the origin of the in situ component of the stellar halo. We found that all the M31 and MW galaxies experienced $10-40$ mergers with dwarf galaxies in a stellar mass range of $10^6-2\times 10^{10}$~\Msun; however, only on to four mergers~(for different galaxies) are significant, where the stellar mass ratio is $0.2-1$~(relative the host at the time of accretion) and all of the mergers~(with a single exception) happened $\approx 7-12$ Gyr ago. Our conclusions are the following.

\begin{itemize}
    \item We found a certain correlation between close pericentric passages of massive satellites and mergers with the evolution of the star formation rate in the host galaxies. In particular, in most of the cases, significant peaks in the SFH correlate with the mergers; however, not all the mergers have the corresponding bursts of star formation in the host galaxy. This likely depends on the parameters of the interaction (orbit, relative mass) and the amount of gas available inside the host.

    \item The mergers induce heating of the in situ stars where the individual events are imprinted as the increase in the vertical motions of stars and the increase in the stellar velocity dispersion for the populations that existed in the disc prior to a given merger. We suggest that the latest significant merger makes the heating effect of the previous ones vanish, resulting in a flattening of the age-velocity dispersion relation for older stars.

    \item All the HESTIA galaxies clearly show the impact of the most massive mergers seen as a sharp increase in the eccentricity~(and decrease in the $V_\phi$) for all the stars that existed in the main progenitor before the merger thus nicely reproducing the Splash-/Plume-like stellar populations recently discovered in the MW using \Gaia data. Moreover, each HESTIA galaxy shows a few kinematically distinct~(seen in the eccentricity and $V_\phi$ distributions) populations caused by different mergers and about $7-14\%$ of these stars have negative angular momentum. 

    \item We have analyse the emergence of the in situ component of the stellar halo, defined kinematically using the Toomre diagram. We have shown that a substantial fraction~($30-40\%$) of the inner stellar halo is formed in situ; while the in situ mass fraction does not exceed $5\%$ of the total stellar halo mass in the outer parts~(out to $20-50$~kpc) of the HESTIA galaxies. The in situ stellar haloes defined kinematically have a broad range of ages with the mean values of $7-9$~Gyr, which is between $1-2$~Gyr younger than the accreted halo component. Although the age distributions of the in situ stellar halo show a number of prominent peaks correlating with the most significant merger events, its substantial mass emerges in between or well after the mergers.
    
    \item We have shown that the diagonal wedges in the $\rm R_{max}-Z_{max}$ distributions can be linked to different merger events. In particular, the lowest angle~($\rm \arctan{Z_{max}/R_{max}}$) wedges correspond to the stars formed after the last significant merger, while the wedges with larger angles are made of stars formed in between the previous significant mergers.

\end{itemize}

In this work, we showed that mergers play a significant role in the evolution of the HESTIA galaxies, shaping the structure and properties of their in situ stellar populations. The collision and subsequent merging of smaller galaxies lead to the enhanced formation of new stars and the redistribution of pre-existing stellar populations that have been found in the MW thanks to the \Gaia data and ongoing large-scale spectroscopic surveys.

\begin{acknowledgements}
We thank the anonymous referee for their valuable comments. SK acknowledgements the HESTIA collaboration for providing access to the simulations. FAG acknowledges support from ANID FONDECYT Regular 1211370 and from the ANID BASAL project FB210003. FAG,  acknowledge funding from the Max Planck Society through a Partner Group grant.  AK is supported by the Ministerio de Ciencia e Innovaci\'{o}n (MICINN) under research grant PID2021-122603NB-C21 and further thanks Field Mice for Emma's house. JS acknowledges support from the French Agence Nationale de la Recherche for the LOCALIZATION project under grant agreements ANR-21-CE31-0019. YH has been partially supported by the Israel Science Foundation grant ISF 1358/18. ET acknowledges support by ETAg grant PRG1006 and by EU through the ERDF CoE TK133.

\newline
{\it Software:} \texttt{IPython} \citep{2007CSE.....9c..21P}, \texttt{Astropy} \citep{2013A&A...558A..33A, 2018AJ....156..123A}, \texttt{NumPy} \citep{2011CSE....13b..22V}, \texttt{SciPy} \citep{2020SciPy-NMeth}, \texttt{Matplotlib} \citep{2007CSE.....9...90H}, \texttt{Pandas} \citep{mckinney-proc-scipy-2010}.
\end{acknowledgements}

\bibliographystyle{aa}
\bibliography{references-1}

\end{document}